# Trends in Workplace Wearable Technologies and Connected-Worker Solutions for Next-Generation Occupational Safety, Health, and Productivity

*Vishal Patel, Austin Chesmore, Christopher M. Legner, and Santosh Pandey\**

The workplace influences the safety, health, and productivity of workers at multiple levels. To protect and promote total worker health, smart hardware, and software tools have emerged for the identification, elimination, substitution, and control of occupational hazards. Wearable devices enable constant monitoring of individual workers and the environment, whereas connected worker solutions provide contextual information and decision support. Here, the recent trends in commercial workplace technologies to monitor and manage occupational risks, injuries, accidents, and diseases are reviewed. Workplace safety wearables for safe lifting, ergonomics, hazard identification, sleep monitoring, fatigue management, and heat and cold stress are discussed. Examples of workplace productivity wearables for asset tracking, augmented reality, gesture and motion control, brain wave sensing, and work stress management are given. Workplace health wearables designed for work-related musculoskeletal disorders, functional movement disorders, respiratory hazards, cardiovascular health, outdoor sun exposure, and continuous glucose monitoring are shown. Connected worker platforms are discussed with information about the architecture, system modules, intelligent operations, and industry applications. Predictive analytics provide contextual information about occupational safety risks, resource allocation, equipment failure, and predictive maintenance. Altogether, these examples highlight the ground-level benefits of real-time visibility about frontline workers, work environment, distributed assets, workforce efficiency, and safety compliance.

## 1. Introduction

The multitude of workplaces come with their inherent distractions, risks, and dangers. Timely identification and remediation of the potential risk factors is critical for the safety, health, and productivity of workers. For this purpose, employers' resort to key enabling technologies (KETs) to monitor, manage, and optimize their workplace assets (e.g., tools, machines, inventory, safety equipment, supporting technologies, and workers). On one hand, workplace wearables (i.e., smart, on-body accessories, and personal protective equipment [PPE]) track the activities, behavior and body status of individual workers. On the other hand, connected worker solutions (i.e., intelligent computing, data analytics, and storage platforms) serve as the central hub for extracting contextual information from distributed networks of workplace wearables for better workflow organization, asset management, and predictive maintenance.

Currently, there are niche applications for workplace wearables and connected worker solutions (herein referred to as workplace technologies) in almost every industry, be it in agriculture, construction, mining, production, healthcare, retail, warehousing, technology, transportation, or automotive industry. For example, workplace technologies are being used to detect awkward work postures, forceful exertions, vibrations, repetitive tasks, physical fatigue, mental acuity and stress, mood and emotions, safety compliance, and rest breaks.[1–5] By building a smart and connected workplace with human-in-a-loop models, workers today are better equipped with situational awareness, field visibility, and remote supervision.[6–9] Previous review articles have summarized smart and intelligent wearables on selected topics and products related to workplace safety, health, and productivity.[1–9] Most of the information on latest commercial products and solutions is spread across different websites, news outlets, insight articles, survey reports and social media.[10–12] This may make it challenging for interested readers and potential adopters to assimilate the information in a digestible format and keep abreast with the emerging workplace technologies, their applications in different worksites, and potential benefits to the workforce and employers.

Our objective here is to provide a comprehensive review of commercial wearables and connected worker solutions for occupational safety, health, and productivity (**Figure 1**). Our intent is

V. Patel, A. Chesmore, C. M. Legner, S. Pandey
Department of Electrical & Computer Engineering
Iowa State University
2126 Coover Hall, Ames, IA 50011, USA
E-mail: pandey@iastate.edu

The ORCID identification number(s) for the author(s) of this article can be found under https://doi.org/10.1002/aisy.202100099.

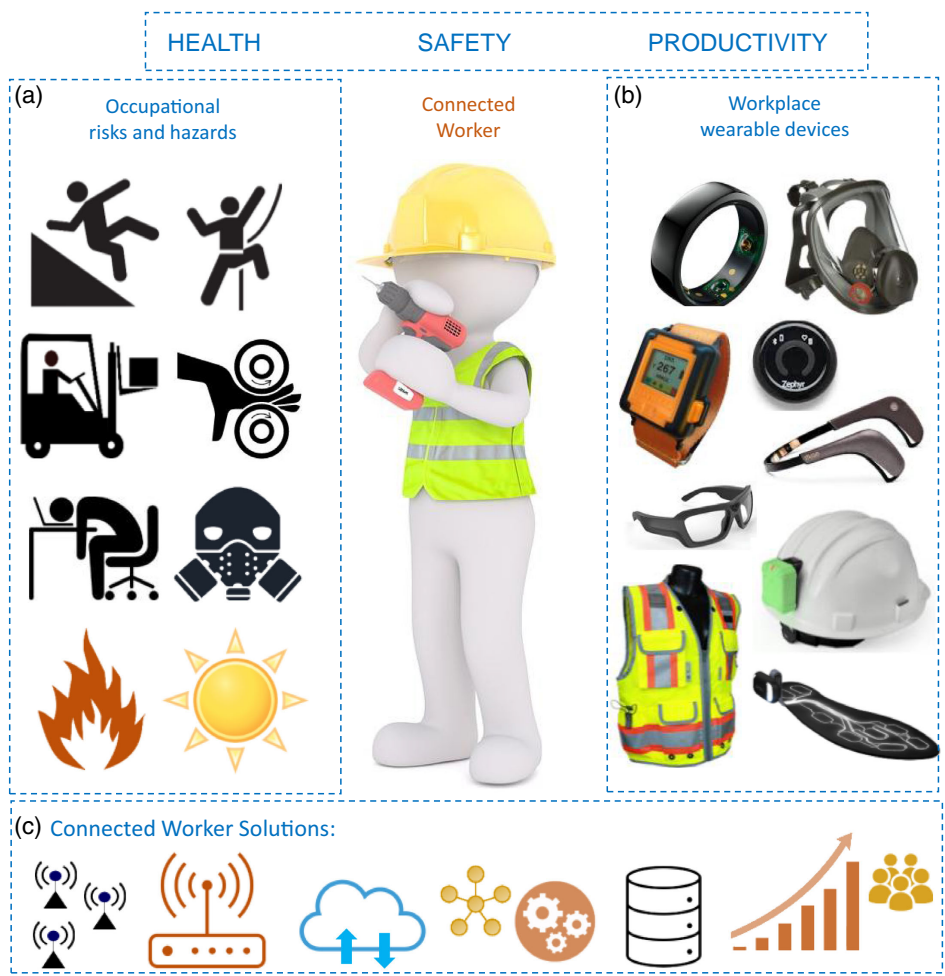

**Figure 1.** Overview of the three topics covered in this review: occupational risks and hazards, workplace wearables, and connected worker solutions. Occupational risks and hazards come from various sources and can adversely affect the workers' safety, productivity, and health. Workplace wearables are smart devices worn on the body as accessories or garments to collect data on physiological and environmental conditions. Connected worker solutions engage the workplace wearables to a centralized hub for data storage, management, analytics, visualization, security, and actions. The vector icons on occupational risks and hazards are reproduced (adapted) with permission from the OSHA website on "Top 10 Most Frequently Cited Standards for Fiscal Year 2020".[16] The product images in (b) are reproduced with permission: smart ring:[229] Copyright 2021, Oura; full face respirator:[63] Copyright 2021, 3M; smart watch:[62] Copyright 2021, Reactec; chest device:[81] Copyright 2021, Zephyr; headband:[114] Copyright 2021, Muse; smart glasses:[231] Copyright 2021, ViewPoint System; smart PPE vest:[63] Copyright 2021, 3M; smart helmet:[61] Copyright 2021, Triax; and smart shoe insoles:[67] Copyright 2021, Arion.

to cover workplace technologies that are already in use or have the potential for use at different worksites. This organization of the article is as follows: Section 2 discusses the occupational risks in different industries and the associated risk-mitigating interventions. Section 3 presents wearables for monitoring occupational safety in the context of heavy lifting, ergonomics, lift assists, danger awareness, hazard identification, fatigue risk management, and extreme temperature stress management. Section 4 presents wearables for boosting worker productivity through the asset tracking, social behavior monitoring, augmented reality (AR), virtual reality (VR), gesture and motion controls, brain-wave sensing, and stress management. Section 5 covers wearables for monitoring occupational health, including musculoskeletal disorders, movement disorders, pulmonary function, cardiovascular health, outdoor sun exposure, thermal comfort, and continuous glucose monitoring (CGM).

Section 6 presents examples of connected worker architectures and product solutions from different vendors. Section 7 discusses the limitations of current workplace technologies and points to consider before investing in them.

## 2. Occupational Risks and Risk-Mitigating Interventions

Occupational injuries, accidents, and diseases have a major impact on the physical, mental, and economic well-being of workers and their families. The International Labor Organization (ILO) estimates that occupational accidents and work-related diseases result in ≈2.78 million deaths and 374 million nonfatal injuries every year.[13] The lost work days worldwide account for nearly 4% of the global gross domestic product (GDP). Workplace incidents tarnish the reputation of companies and cause significant financial burden (e.g., medical insurance premiums, worker compensation, litigation costs, lost productivity, and the costs of hiring and training replacements).[13,14] One estimate suggests that workplace injuries cost USA companies more than one billion dollars every week.[15]

**Table 1** shows the occupational risks in different industries, the workers at increased risk, the risk-mitigating interventions, and PPE commonly used to safeguard workers as compiled from safety and health reports.[13–16] Specific work groups are at higher risks for certain injuries, infections, and diseases.[14] For instance, healthcare workers are prone to fatigue, overexertion, and hospital infection. Office workers are susceptible to neck and back pain, chemical factory workers have greater risks of respiratory illnesses, and rubber industry workers face an increased risk of stomach cancer. Hearing loss is commonly associated with workers in the mining, construction, manufacturing, and entertainment industry. Asbestos is a risk factor for mesothelioma and ovarian cancer, plutonium and vinyl chloride are risk factors for liver cancer, and trichloroethylene (a common industrial solvent) is a risk factor for kidney cancer.[17–20] Intestinal and parasitic infections are prevalent in workers exposed to unhealthy, pathogen-borne environments, such as those in the military, agricultural and animal farms, construction sites, and wastewater treatment facilities.[21–25]

To facilitate total worker health (TWH), effective risk-mitigating interventions are constantly needed through close collaboration among safety organizations, government agencies, employers, and workers.[26–28] The key organizations responsible for formulating, implementing, protecting, and promoting the occupational safety and health guidelines are the International Labor Organization (ILO), World Health Organization (WHO), Environmental Protection Agency (EPA), Occupational Safety and Health Administration (OSHA), National Institute for Occupational Safety and Health (NIOSH), Occupational Safety and Health Review Commission, and Centers for Disease Control and Prevention (CDC). The important labor laws and standards defining the occupational safety and health guidelines are the USA. Occupational Safety and Health Act of 1970, the ILO's International Labor Standards of 1919, Canada's Occupational Health and Safety Act of 1979, the UK. Health and Safety at Work Act of 1974, the USA. Mine Safety and Health Act of 1977, and Australia's Work Health and Safety Act of 2011.[29–31] The OSHA Whistleblower Protection Program protects workers from retaliation by employers when they report workplace violations.[16]

Table 1. Occupational risks in different work environments, workers at increased risk, risk-mitigating interventions, and PPE.

| Occupational risks | Workers at increased risk | Risk-mitigating interventions | PPE |
|---|---|---|---|
| Asthma, COPD, respiratory infections, and diseases | Workers with exposure to allergens and irritants | Ventilation, pulmonary function testing, exposure monitoring, emission containment | Dust masks, respirators, gas masks |
| Behavioral, mental, or neurological disorders | Workers in any work environment | Identification of distress, drug use, and mental illness; healthy work–life balance | Trackers to monitor and manage stress and physical activity |
| Cardiovascular diseases, ischemic heart disease, stroke | Workers with exposure to lead, pollutants or high stress environments | Lead exposure monitoring and reduction; stress-reducing lifestyle | Safety glasses, face shields, dust masks, respirators |
| Electrical hazard, electric shock, electrocution | Electricians; Workers in manufacturing, construction, agriculture and power plants | Ground-fault protection, electrical standards, hazard assessment, employee training, PPE | Safety boots, flame-resistant clothing, face shields, hard hats |
| Exertion, physical inactivity, overweight, obesity | Workers in any work environment, particularly in offices and assembly plants | Ergonomics, physical activity and yoga, healthy diet, lifestyle monitoring | Belts, harness, posture correctors, activity monitoring devices |
| Eye injuries, cataract | Outdoor workers, welders, carpenters; Workers with high UV exposure | Safety regulations, eye protection policies, eye wash stations, risk assessment, PPE | Safety glasses, face shields, goggles, welding helmets |
| Falls, slips, trips | Workers in manufacturing, agriculture, construction, fishery and mining | Window guards, grab rails, lifelines, safety nets, lighting, equipment inspection, PPE | Harnesses, hard hats, non-slip shoes, fall-arrest systems |
| Fire, heat, hot substances | Fire-fighters, first responders; Workers in mining, manufacturing and construction | Safety regulations, smoke detectors, fire suppression systems, evacuation plans, PPE | Face shields, insulated clothing, fire resistant clothing, helmets |
| Hearing loss | Workers in mining, manufacturing, construction and entertainment industry | Limiting noise exposure, using noise-reducing controls, PPE | Earplugs, hearing aids, hearing protectors |
| Injuries from motor vehicles, mechanical forces, or road traffic | Workers operating heavy machinery (e.g., in construction and industrial plants) | Backup cameras, spotters, proximity detectors, driving laws, vehicle safety, drug testing | High visibility safety vests, hard hats, safety lights |
| Interpersonal violence | Cashiers, teachers, policemen | Regulation, employee training, PPE | Kevlar vests, body armor, riot gear |
| Mesothelioma, melanoma, mouth or lung cancer | Industrial, agricultural and other workers exposed to carcinogens or radiation | Exposure monitoring and reduction; capsulation and closed processes | Safety glasses, masks, face shields |
| Musculoskeletal disorders | Workers in agriculture, forestry, fishery, production and service industry | Ergonomics, specialized equipment and tools, adequate working hours, risk assessment | Exoskeletons, braces, belts |
| Neonatal and congenital conditions | Workers with exposure to certain chemicals and second-hand smoke | Elimination of reproductive risks, engineering controls, no-smoke policies | Dust masks, respirators, gas masks, clean suits, gloves |
| Repetitive motion with microtasks | Workers in any work environment | Ergonomics, regular breaks, monitoring physical activity and lifestyle choices | Posture-correcting harnesses |
| Skin diseases (e.g., contact dermatitis) | Hairdressers, cosmetologists, cleaners, painters, healthcare workers | Exposure monitoring; reduction of allergens or irritants; smoke-free policies, PPE | Gloves, face masks, specialized clothing and suits |
| Water-related incidents, drowning | Fishermen, first responders; Workers in boats, ships, ferry | Safety rules, guard rails, evacuation plans | Life jackets, buoyant work vests, safety glasses, flotation devices |

The employers, particularly the corporate leadership and management staff, are ultimately responsible for implementing and adopting the risk-mitigating interventions to ensure a safe and healthy work climate.[32] For outdoor work activities, employers may provide standard PPE to their labor workforce, such as steel toe cap shoes, safety vests, harnesses, hard hats, earmuffs, and goggles. In this regard, workplace technologies with real-time data collection and predictive analytics are worthwhile investments to assess the internal and external workloads of workers using quantifiable metrics and measurable outcomes.[33] In addition, employers may offer a variety of perks and rewards to positively influence the work-life balance of their employees. Workplace wellness programs play a vital role in supporting healthy choices and improving the quality of life of workers to benefit the employers in the long term.[34] Behavioral and economic incentives (such as fitness and yoga instruction, stress management classes, healthy food canteens, transit options, employee assistance programs, and worksite health clinics) have shown to boost the workforce morale.[34] In the case of injuries and accidents, employers may offer workers compensation benefits, temporary or permanent total disability benefits, social security disability insurance, and death benefits. To raise global awareness about the prevention of occupational injuries, accidents and diseases, the ILO World Day for Safety and Health is observed annually on April 28.[35] The 2020 campaign theme was the outbreak of infectious diseases at work, and in particular on how to combat the COVID-19 pandemic and resume work activities in a sustainable manner.

## 3. Wearables for Occupational Safety Monitoring

In this section, the following categories of wearable devices for monitoring occupational safety are discussed: i) heavy lifting, ergonomics and lift assists, ii) danger awareness, hazard identification, and fatigue risk management, and iii) heat and cold stress detection. **Table 2** shows the commercial devices for monitoring worker safety, companies that produce them, their classification, components, and safety applications. **Figure 2** shows examples of wearable technology products for monitoring occupational safety.

### 3.1. Heavy Lifting, Ergonomics, and Lift Assists

Lifting heavy objects can lead to back sprains, muscle pulls, and injuries in the wrist, elbow, or spinal cord.[36–38] Workers encounter heavy lifting activities in several workplaces, including those in construction, manufacturing, warehousing, retail, and transportation. Several agricultural activities from plantation to harvesting require heavy lifting and repetitive tasks.[39] Within the healthcare service industry, emergency medical services (EMS) workers routinely encounter situations in which they must lift prehospitalized patients, resulting in overexertion and bodily reactions (i.e., sprains and strains). According to the NIOSH, heavy lifting is the most common cause of occupational injuries for these EMS workers.[40]

Wearables for ergonomic hazards can detect awkward working postures during heavy lifting (e.g., awkward bending, reaching, gripping, and repetitive lifting events) and provide real-time safety alerts to correct the wearer's posture.[41] For example, the kinetic reflex is a smart wearable device worn on a belt or waistband to detect high-risk postures in industrial workers.[42] The kinetic device uses motion sensors and biomechanical analysis to provide real-time vibrational feedback. The kinetic reflex wearable and software analytics platform is used for injury reduction at the workplace by automatically detecting unsafe biomechanics at work (e.g., bending, twisting, and overreaching) that could eventually lead to injury. A continuous coaching with light vibration is provided to workers, along with alerts, goals, and rewards to create and sustain new behavior or habits. Kinetic estimates that its technology can reduce potentially unsafe postures by 84%, thereby preventing lifting injuries commonly encountered in shipping, packaging, and last mile delivery of goods. A report by Perr&Knight, a top USA actuarial consulting firm, found that the use of kinetic reflex wearable technology resulted in 50–60% reduction of injury frequency (i.e., frequency of strain and sprain claims) and 72% reduction of lost work days.

Lift assists, such as exoskeletons and exosuits, have proved vital for heavy and repetitive lifting encountered in industrial jobs. Exoskeletons act as mechanical extensions of the wearer to assist in a range of motions while providing ergonomic benefit and reduced muscle strain.[43,44] The following are examples of exoskeletons designed for upper body ergonomics:

The StrongArm ErgoSkeleton lift assist system is a workplace safety vest designed for the "industrial athlete."[45] The StrongArm fuse platform uses a risk monitoring device to collect data on the body sagittal motion, twist motion, and lateral motion.[46] Machine learning (ML) algorithms evaluate safety risks, and alert the user via haptic feedback. The StrongArm ErgoSkeleton is used to reduce arm fatigue, transfer the load from the upper body to the legs, reduce strain and injuries to the lower back, and inculcate proper lifting techniques at the workplace. A safety score is maintained for every worker to provide a holistic view of the safety situation for every facility, job shift, and job function.

The Levitate Technologies AirFrame wearable harness is used to fight fatigue in workers' upper extremities and improve upper extremity musculoskeletal health, especially in jobs that require repetitive arm movement or static elevation of the arms.[47] The wearable harness transfers the weight of the arms from the shoulders and upper body to the hips and body core, thereby reducing stress and injury to the arm. The AirFrame is automatically activated as the arm is raised and is deactivated as the arm is lowered. The harness is purely mechanical and operates by a system of pulleys and cables, thus requiring no electrical power. In a significant milestone for the exoskeleton industry, the Toyota's Woodstock plant has made the AirFrame suit a mandatory PPE for its workers in the welding shop, recognizing the risks of working overhead as a key contributor to musculoskeletal injuries.

Table 2. Examples of wearable devices to monitor the safety of workers.

| Company | Sample products | Classification | Components | Safety applications |
| --- | --- | --- | --- | --- |
| 3M | 3M DBI-SALA ExoFit NEX, ExoFit STRATA, Nano-Lok | Vest, harness, helmets, PPE | Full-body harnesses, lifeline systems, anchorage connectors | Fall protection, machine operations, assembly, demolition, remediation, electrical jobs |
| Bioservo | IronMan System, Remote Control, Battery Pack, App | Glove | Pressure sensors in five fingers embedded processor, actuator, battery | Strengthen gripping and grasping action; adjust applied force and balance between fingers |
| Equivital | EQ02 LifeMonitor, Black Ghost monitoring system | Chest Device | Sensors for activity, posture, heart rate, breathing rate, core body temperature | Mobile human monitoring; 24/7 health and safety observations in extreme environments |
| GuardHat | Hard hat, Tag, Mobile App | Helmet | Sensors for noise, gas, temperature, and pressure; integration to external sensors | Real-time situational awareness and location monitor; audio and video features |
| Honeywell | BioHarness, RAELink3, ProRAE App | Chest Harness | Sensors for heart rate, body temperature, breathing rate, posture | Real-time measurement of vitals; knowledge of location and health status; manage safety |
| Kenzen | Kenzen Patch, Kenzen Monitor, App | Patch | Biosensors for heart rate, sweat, temperature, activity | Real-time reports of heat stress indicators; worksite heat assessment; training and consultation |
| Kinetic | Reflex | Belt device | Inertial measurement units, alert system | Detect high-risk bending, twisting and reaching |
| Levitate | AirFrame | Exoskeleton | Mechanical support system, driven by pulleys, no electrical power | Distribute upper body weight; reduce fatigue and muscle stress |
| Modjoul | Modjoul SmartBelt | Belt | Eight sensors, vibration alert response sensor, GPS, ABS belt buckle | Collects data on location, activity and environment; identify risky activities; haptic response |
| Reactec | HAVwear, Docking Station, Reactec Analytics | Wrist monitor | Sensors for hand arm vibration (HAV), whole body vibration, noise, gas, dust | Hand Arm Vibration (HAV) risk assessment; tool testing; exposure monitoring; risk control appraisal |
| Realwear | HMT-1 | Headset | motion sensors, GPS, digital microphones, 16 MP camera, 4-axis image stabilization | Hands-free digital workflow, virtual assistance, remote mentoring, data visualization |
| Ripple Safety | Ripple Button, Safety App, Monitoring Service | Safety button | Access to mobile phone features, Bluetooth connectivity, lithium batteries | Send alerts and location data, press unique buttons for non-emergency or emergency response |
| Scan-Link | Scan-Link Armor RFID System, Safety Apparel | RFID system | RFID tags, display, external alarm, antenna (865 to 927 MHz, 20 feet range) | Alert the driver about the presence of ground workers; prevent collisions and personal injury |
| SmartCap | LifeBand, Life App | Flexible band | EEG sensors, brainwave technology detectors, Bluetooth communication | Measure fatigue and alertness at real-time; eliminate microsleeps; support wellness |
| SolePower | SmartBoots | Shoes | Force and motion sensors, control boards | Measure gait changes, impact and fatigue |
| StrongArm | ErgoSkeleton Lift Assist, FUSE Risk Management | Harness | Motion sensors, ML analytics, posture feedback system | Measure risk of musculoskeletal injuries, give real-time safety score, follows OSHA lifting guidelines |
| Triax | Spot-r Clip, Tags | Clip | Three-axis accelerometer, gyroscope, altimeter | Fall alerts; self-alerts and logging |
| Zephyr & Medtronic | Zephyr Performance, BioModule device | Chest device | Sensors for heart rate, breathing, activity, posture, accelerometry, impact | Track body vitals, biomechanical metrics, heat stress risks, physical load monitoring |

Paexo from Ottobock and Skelex 360 from Skelex have been tested by Audi, the German automaker, to assist with overhead and above-shoulder jobs in its assembly, toolmaking and paint shops at the Ingolstadt site.[48] These exoskeletons are designed as backpacks and have arm shell structures that absorb some of the weight of the arms and redistribute it to the hips and thighs of the wearer. The exoskeletons help to reduce physical strain and protect the musculoskeletal system during a variety of overhead or above-shoulder activities, such as electrical or mechanical installations, structural assembly, engine repairs, welding of metallic framework, riveting activities maintenance work, pretreatment/sanding/paint coating, foil application etc.

The Noonee Chairless Chair is an exoskeleton worn at the back of the legs by assembly line workers.[49] The exoskeleton structure has surfaces that support the buttocks and thighs, and two struts that contour to the shape of the legs. The Noonee Chairless Chair allows the user to sit dynamically and switch effortlessly between sitting, standing, and walking activities in industrial production, picking, and assembly lines. This promotes ergonomic posture and relieves strain from the legs, knees, and ankles.

The Sarcos Robotic Guardian XO is a full-body exoskeleton that assists with repetitive lifting, heavy lifting, and long durations of physical activity.[50] Some use cases of the Sarcos Guardian XO in assembly and manufacturing (e.g., unload boxes and crates, manipulate parts and heavy tools, welding, and kitting industrial equipment), construction (e.g., moving, installing, dismantling, and lifting heavy tools and equipment), and aviation (e.g., minimize risks of stress/strain from physically demanding jobs). The Sarcos Guardian XO suit amplifies the operator strength up to 20 times (with a maximum payload of 200 pounds). The SuitX MAX flexible exoskeleton series is specially adapted to reduce the muscle force of the lower back, shoulder, and legs.[51] They are designed to reduce muscle strain, reduce risk of injury, and improve the quality of life. There are three modules (backX, shoulderX, and legX) that can function independently or in unison to reduce the strain on body joints. The shoulderX module reduces shoulder fatigue when

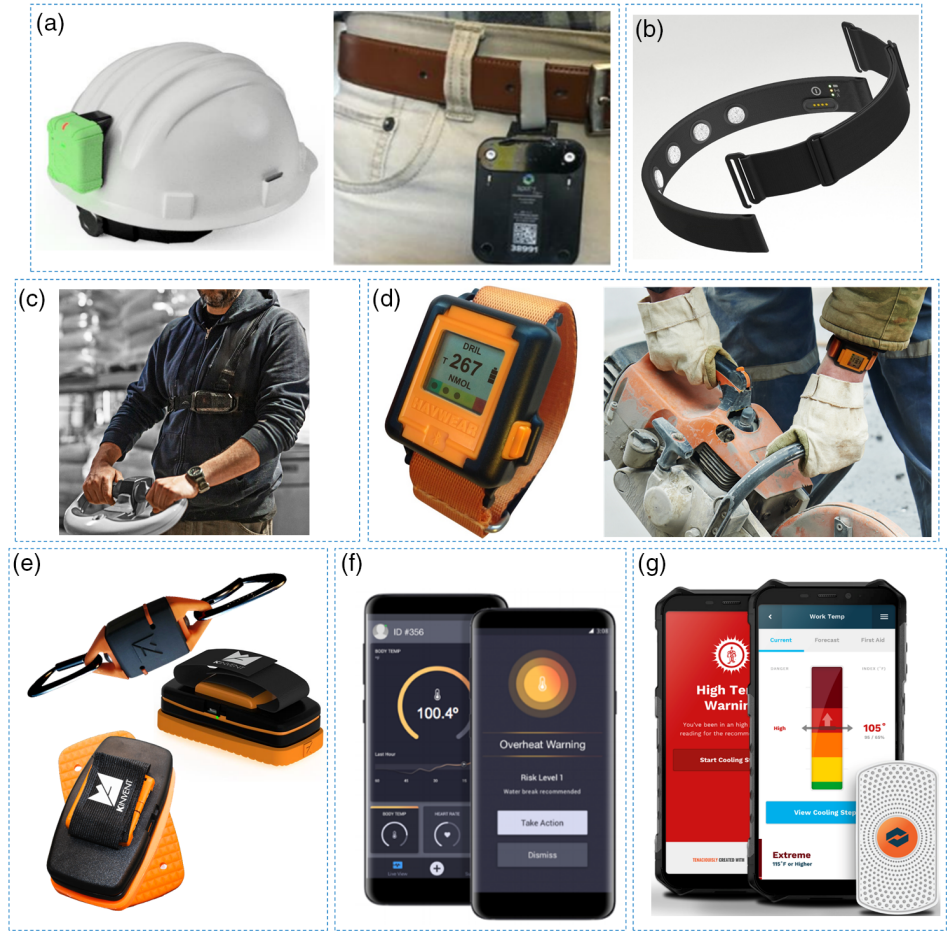

**Figure 2.** Examples of wearables for occupational safety monitoring. a) The Triax Proximity Trace consists of TraceTags affixed to hard hats for monitoring safe working distance and interactions. b) The SmartCap LifeBand measures brain waves to estimate alertness and fatigue levels at work. c) The StrongArm ErgoSkeleton Lift Assist helps with heavy lifting by transferring the load from the upper body and back to the legs. d) Reactec HAVwear is designed for the risk management of hand arm vibrations and provides real-time safety analytics and alerts. e) K-invent dynamometers consist of force transducers with biofeedback to evaluate muscle strength and endurance. f) Kenzen's personalized app and smart PPE enable the continuous safety monitoring of workers by tracking the microclimate, core body temperature, heart rate and exertion. g) The Corvex Temperature Management platform monitors the work temperature of frontline workers, providing them with situational awareness and data transparency. The figures were reproduced with permission: a) Reproduced with permission.[61] Copyright 2021, Triax; b) Reproduced with permission.[118] Copyright 2021, SmartCap; c) Reproduced with permission.[45] Copyright 2021, StrongArm; d) Reproduced with permission.[62] Copyright 2021, Reactec; e) Reproduced with permission.[139] Copyright 2021, K-invent; f) Reproduced with permission.[82] Copyright 2021, Kenzen; and g) Reproduced with permission.[230] Copyright 2021, Corvex.

arms are raised during above-shoulder work, the backX module reduces back fatigue during a bending position, and the legX module reduces thigh fatigue during a squatted position. The following are examples of exoskeletons designed for hand ergonomics:

The Bioservo Ironhand soft exoskeleton glove enhances the gripping strength of the hand while using less grip force. An embedded processor calculates the needed force, and then the artificial tendons apply the extra force.[52] The glove consists of sensors in all five fingers that adjust the applied force and balance between fingers. This feature reduces strain injuries and fatigue during tasks involving intensive grasping actions. The glove is always connected by 4G using an in-built sim card or WiFi. The battery life is ≈6–8 h and the weight is around 6 pounds (or 2.75 kg). The Bioservo Ironhand is used as a preventive ergonomic measure to reduce incidents of hand-related repetitive strain injuries (RSIs) at the workplace. Some use cases of the Bioservo Ironhand are in preassembly (e.g., manual maneuvering of the hoist), assembly (e.g., pressing in, clipping, and compressing), logistics (e.g., lifting and carrying off parts), warehousing (e.g., lifting, holding, and carrying boxes), and construction (e.g., grip-intensive and repetitive tasks).

The RoboGlove is a robotic grasping wearable assist technology developed by the National Aeronautics and Space Administration (NASA) and General Motors. It is used to help assembly line operators working long hours and performing repetitive tasks, and could be applied in a variety of markets from manufacturing to medical rehabilitation.[53] The RoboGlove reduces the gripping force needed while operating a tool and has pressure sensors at the fingertips to detect when an object is ready to be grasped. Synthetic tendons then retract to bring the RoboGlove fingers into a gripping position. The object is released when the gripping action is relaxed. The RoboGlove is now licensed to Bioservo that has developed its own Soft Extra Muscle Glove.[54]

## 3.2. Danger Awareness, Hazard Identification, and Fatigue Risk Management

Workers can be subject to unpredicted dangers and hazards, such as when bypassing safeguards during heavy machinery operations.[55,56] Supervisors need to ensure that workers are immediately aware of potential dangers to avoid accidents.[57] Using wearables, workers can inform their supervisors about their location, fatigue levels, health status, and the surrounding environment. This digital connectivity and data transparency allow supervisors to remotely observe workers, check for safety compliance, assess potential dangers, and send timely alerts or requests for help. The following are examples of workplace wearables for danger awareness, hazard identification, and accident prevention:

Guardhat integrates sensors into a smart hard hat (certified ANSI.Z89.14 Type I Class-G industrial hardhat) to monitor multiple parameters, such as altitude, proximity, falls, noise, gas, temperature, and pressure.[58] Guardhat has different multimedia options (light emitting diodes (LEDs), audio speakers, microphones, 13MP camera, push-to-talk (PTT), voice over internet protocol (VoIP), and Video Uplink), maximum weight (with suspension) of 800 gm, battery capacity of 4800 mAh (typically 8-12 h), and a 3D location accuracy of up to 1 m. Its Real-time Location System has the ability to locate assets, personnel, and sensors with submeter resolution to alert onsite personnel of operating equipment and hazardous or restricted zones. Real-time indoor and outdoor location tracking is performed using the Global Positioning System (GPS)/global navigation satellite system (GNSS) technology. Asset identification, proximity detection, and presence detection are enabled using Bluetooth Low Energy (BLE) beacons and radio frequency identification (RFID) technology. The Real-time Situational Awareness System provides situational context to human–machine interactions to make immediate decisions, prevent accidents, and improve critical response metrics. The Guardhat is used to detect proximity to hazardous locations and moving equipment, provide early warning for potential exposure to hazmat or environmental conditions, monitor the lone worker or the driver/machine operator remotely, monitor access control at restricted areas (e.g., lockouts, geofencing, zone marking etc.), check and remind workers for wearing the correct PPE, and provide emergency response or evacuation notifications in a timely manner.

The Scan-Link Armor RFID System is designed to be attached to safety vests and hard hats.[59] The Scan-Link system consists of an antenna unit, display, beeper alarm, mounting bracket, and safety apparel. Its RFID tags in the antenna unit is used to alert the mobile equipment operator (e.g., drivers and ground workers) about the presence of persons (e.g., other ground workers, contractors, and visitors) and danger zones in the construction site within a wide range from the operator (25–30 feet behind and 13 feet across). The antenna unit records the date/time and apparel ID of every detection event, and this information is used by the system to self-learn and reduce false alarms. The available frequencies for the antenna unit are 902.3–927.7, 920–926, and 865.7–867.5 MHz. One unique feature of the system is that, rather than detecting objects (e.g., gravel, sand piles, safety cones, and safety barriers), the Scan-Link RFID tags can detect and differentiate only persons in the vicinity, alerting the mobile operator to exercise caution. The Scan-Link technologies have been used in various industries (e.g., oil and gas, construction, waste, forestry, aggregate, and municipal) to increase safety practices and be compliant of safety regulations at the workplace.

The Honeywell BioHarness physiological monitoring system supports wearable sensors that measure users' heart rate (0–240 beats per minute), breath rate (0–120 breaths per minute), temperature (10–60 °C), activity (±16 Vector Magnitude Units), posture (0 to 180 degrees), and location.[60] Its RAELink3 is a portable wireless transmitter that allows real-time data viewing within a 2 mile radius, whereas its ProRAE Guardian software provides continuous information on the location and health status of remote workers. Its RFID tags can be attached to any Honeywell safety equipment to track its location, its usage, and maintenance performed by field personnel such as firefighters, first responders or factory workers. The Honeywell BioHarness provides a quick assessment of worker's health status on the field so that work levels can be modified to mitigate potential health risks (such as heat stress) from hazardous atmospheres.

The Triax Technologies Spot-r platform provides real-time visibility on the location, utilization, and safety of both workers and equipment.[61] The Spot-r Clip is a clip-on device worn by workers to automatically transmit information about their location (e.g., floor and shop), attendance (e.g., check-in and check-out), and safety compliance. The clip-on device automatically detects falls, informs supervisors about the nature of falls, and calls for emergency help through the Spot-r network. The Spot-r EquipTag provides information about the equipment (e.g., its location, operator, and usage) to its network to manage idle time, the number of hours logged in, and operational use. The Spot-r EquipTags, POI tags, and dashboard have been used in oil and gas refineries, construction sites, and heavy industrial facilities for site visibility (active and occupied sites), worker/site safety (high volume traffic monitoring, lone worker check-ins, and confined spaces), site security (restricted area monitoring), and automated identification of productivity gains (productive versus unproductive areas, and workplace bottlenecks) at the work site

Reactec has developed a managed service solution for the exposure to Hand Arm Vibrations (HAV) while operating machinery.[62] Its Razor device captures data from multiple risk exposure sensors (e.g., vibration, noise, gas, and dust) housed in its HAVwear watches. The Havwear watch has a 17 mm LCD screen, lithium-ion rechargeable battery, 1000 full charging cycles, up to 12–36 h battery life with Bluetooth enabled, memory to store up to 202 data records, and 13.56 MHz data communication module. The HAVCare service conducts HAV risk assessments, exposure monitoring, and risk control appraisals through HAV consultants. The Reactec Analytics Platform provides actionable insights to support lone workers or remote supervisors with sound and vibratory alert signals when the risk exposure thresholds are reached. The Reactec HAVwear is used to track the tool usage, calculate and display daily exposure points, send alerts when the daily exposure values have exceeded, generate reports on worker daily exposure and interventions, record the effectiveness of controls and real-time risk assessment.

3M Technologies is a leading manufacturer of a wide range of smart PPE, safety vests and harnesses, lifeline systems, protective coveralls, protective eyewear, electronic earmuffs, and particulate respirators.[63] Their PPE products are used within the subject areas of fall protection, firefighters and first responder safety, head and face protection, hearing protection and communication, protective eyewear, respiratory protection, welding safety, etc. The 3M DBI-SALA Smart Lock Connected SRL technology is integrated in several PPE gadgets, and uses Bluetooth-enabled sensors for incident reporting, compliance, and inventory management. The 3M Connected PPE system provides actionable insights to workers and managers about the current conditions and historical usage of the PPE gadgets. Together, the 3M Connected Safety Solutions are designed to improve worker safety, safety compliance, and safety process automation.

AerNos has developed a range of smart nano gas sensors used to detect hazardous gases, gas leaks, volatile organic compounds, and air pollutants.[64] Its AerSIP is a $5 \times 5\,mm^2$ System-in-Package (SiP) component that combines a microelectromechanical (MEMS) nanosensing module and a mixed-signal application specific integrated circuit (ASIC) module. Its AerBand is a wearable band used to monitor air pollutants with parts per billion sensitivity and is customizable for specific gases (e.g., formaldehyde, nitrogen oxides, methane, and ozone). The AerBand provides a personal 24/7 monitoring of the air quality, pollutants, and other gases both indoor and outdoor to evaluate the efficacy of pollutants' exposure reduction strategies. The AerNos AerIoT and AerBand Data Cloud Platform & Auto API can be used to store and manage all the recorded data for 24/7 industrial safety monitoring.

To avoid dangerous situations from physical fatigue, a category of smart shoes and insoles has emerged for fatigue risk management that track changes in user's movement, posture and coordination.[65] The following are examples of smart shoes and insoles for the outdoor workplace.

The SolePower Work Boots are designed to monitor the worker's exertion, location and environment in real time.[66] The boots incorporate low power sensors with the capability to send data to the cloud. The mechanical compression of the sole from walking steps is leveraged to power the electronics, and the stored energy can be scavenged in a resource-limited setting. The SolePower boots have been used by industrial workers, military personnel, and first responders to measure fatigue, provide emergency alerts, improve situational awareness, and enhance transparency about the workforce.

The Arion Smart Shoe Insoles and Footpods are used to monitor the wearer's body posture, movement, and running technique to eventually let you run faster, longer, and safer.[67] `The flexible smart insoles are just 2 mm thin, and laden with eight pressure-sensitive sensors to track parameters related to the user's balance, stability, cadence, foot strike, step length, contact time, and flight time. The battery life of the insoles is up to 7 h of run time. The Arion coaching engine provides real-time audio feedback to spread the loading of the body, correct the wearer's posture, and reduce injury. In addition, StryD and RunVi recently launched smart insoles with biomechanical sensors and real-time adaptive coaching to analyze the movement dynamics and avoid injury.

### 3.3. Heat and Cold Stress Detection

Exposure to extreme temperatures is a major concern for outdoor workers, such as emergency responders, police and law enforcement officers, sanitation workers, snow removal crews, and those used in cold storage warehouses or meat packaging plants. The risk factors for hypothermia (body temperature $<95\,°F$) are exposure to cold temperatures, poor physical conditioning, exhaustion, lack of sufficient warm protective clothing, or predisposing health conditions (e.g., diabetes, hypothyroidism, hypertension, trauma, or consumption of drugs or illicit alcohol).[68,69] Hypothermia leads to symptoms such as shivering, slow pulse and breathing, slurred speech, exhaustion, and lack of coordination.[70–72] Prolonged exposure to cold stress can result in frost bite, trench foot, and chilblains.[73–75] On the other hand, the risk factors for hyperthermia (body temperature $>100.4\,°F$) are hot and humid environments, overexertion and excessive workload, thyroid storm, hypothalamic hemorrhage, toxic ingestions, predisposing health conditions (e.g., hypertension, dehydration, being over- or underweight, or consumption of alcohol or illicit drugs). Hyperthermia leads to symptoms such as headache, nausea, fatigue, dehydration, excessive sweating, muscle cramps, rapid breathing, vomiting, and mental confusion. In extreme cases, hyperthermia can result in heat stroke, hallucinations, seizures, and coma.[76–80]

Devices for monitoring the skin temperature include glass thermometers, infrared thermometers, thermal imagers, and wearables. While most wearables are capable of monitoring the skin temperature in a continuous and unobtrusive manner, it is challenging to measure the core body temperature in the same manner. By integrating biometrics and data analytics to estimate core body temperature, employers can remotely assess the heat or cold stress levels of their workers and take the necessary preventive measures. The following are examples of workplace wearables for detecting heat and cold stress.

Medtronic Zephyr has developed the Zephyr Performance System as the premier biometric device for human performance monitoring.[81] Using proprietary technology, Zephyr is able to derive physiological and biomechanical measurements to gain insights with more actionable data to assess performance and safety of an individual or a group. The chest-mounted Zephyr Biomodule collects over 100 data points per second of triaxial accelerometry and samples ECG heart rate from 250 to 1000 Hz. Up to 450 h of data can be collected on up to 100 subjects simultaneously for each Biomodule. The Biomodule can communicate up to 400 yards with the Zephyr Omnisense software and transmit data to a cloud-based server for either in-person or remote surveillance. The Omnisense software can translate this raw data into actionable information such as Heart Rate, Breathing Rate, Heart Rate Variability/ Stress, Estimate Core Body Temperature, Caloric Burn, or Impact Data. The Omnisense Software has over 30 parameters for review, along with customizable and automated reporting functionality. These reports and measurements can be used in a variety of environments and in many ways to achieve greater results, improve safety, and reduce the risk of injury. The Zephyr ECHO communication system transmits up to 200 yards—and up to 400 yards when used with field repeaters. The Zephyr technology has been used by professional and collegiate sports teams, military, NASA, first responders, and for industrial applications such as mining, oil and gas, and construction.

Kenzen offers the Kenzen Patch to monitor physiological parameters associated with heat stress such as heart rate with multi-LED PPG sensor, core body temperature, worker microclimate and sweat rate, body temperature, motion metrics, and activity levels.[82] The Kenzen Monitor is worn on the torso or upper arm and provides regular feedback on vital signs and performance with health insights. The Kenzen Team Dashboard displays data on the entire team. The Kenzen Safety Intelligence generates safety reports based on alerts, actions taken, and flags captured. Kenzen solutions help to predict and prevent heat-related injuries and illnesses by identifying high-risk workers, adjusting shift times, calculating optimal work–rest cycles, allowing workers to improve their health indicators, and training teams to look forward to possible heat-related injuries. Their technology is used to improve worker safety in a variety of workplace settings including nursing, emergency services, oil and gas, refinery, construction, power generation, renewable energy, and mining.

The Equivital EQ LifeMonitor system integrates sensors that measure the wearer's position/movement, activity, oxygen saturation, respiratory rate, and respiratory waveforms, ECG waveforms, heart rate, R–R interval, galvanic skin response, and skin temperature.[83] The 8 Gb memory allows for up to 50 days of continuous logging of physiological data, and the battery life is up to 48 h. The Equivital BlackHost Monitoring System enables mobile human monitoring of the entire team in challenging and hazardous conditions. The Equivital system has been deployed in field studies involving first responders, military personnel, industrial workers, and academic researchers. The Equivital system has been used to track multiple vital signs around the clock with applications in remote health monitoring (e.g., for heat stress prevention, sleep research, sports monitoring, optimized team training, military training, and bio-feedback). It is possible to seamlessly integrate third-party devices such as the Nonin oxygen saturation probes, VitalSense dermal patch, and the VitalSense and BodyCAP core temperature capsules.

North Star BlueScope Steel, a global steel producer and supplier to Australia and North America, embeds different IBM sensors in helmets and wristbands to monitor ground workers' skin body temperature, heart rate, and activity in real time.[84] The IBM's Employee Wellness and Safety Solution uses cognitive computing through its IBM Watson technology to generate customized safety guidelines on when and how to avoid heat stress and exertion, especially if temperatures rise to unsafe limits.

## 4. Wearables for Occupational Productivity Monitoring

In this section, the following categories of wearable devices for monitoring occupational productivity are discussed: i) Asset tracking and social behavior monitoring, ii) AR and VR, iii) gesture and motion control, and iv) mental acuity, brain-wave sensing, and occupational stress management. **Table 3** shows the commercial devices for monitoring worker productivity, the companies that produce them, their classification, components, and applications. **Figure 3** shows examples of wearable technology products for monitoring occupational productivity.

### 4.1. Asset Tracking and Social Behavior Monitoring

Rugged and lightweight wearables are indispensable for tracking assets and the workforce in warehouses, grocery stores, manufacturing sites, factory floors, field stations, and even the battlefield. The following are examples of workplace wearables for tracking physical goods and managing inventory.

The Rufus Ring, Rufus ScanGlove Barcode Scanner, and Rufus Cuff Pro are used to enable faster barcode scanning and inventory tracking.[85] The Rufus Cuff Pro is a wearable computer supported by Android 9 and has a 5.5 in. display with over 12 h battery life. The Rufus WorkHero is the centralized dashboard for productivity analytics to optimize workflows, reduce costs, and lower liabilities. The Rufus WorkHero is used to track parameters such as pickup rates, activity levels, tasks completed, and labor costs. This improves the workers' visibility to warehouse operations and reduces the pick speed and labor costs by up to 55%. The working range for omnidirectional scanning is around 20 feet or 6 m. Third-party logistics software (e.g., Enterprise Resource Planning [ERP], workflow Resource Planning [WRP], legacy systems, and Oracle NetSuites) can seamlessly connect with the Rufus WorkHero using a mobile app.

The Hyco W562 is a rugged smartwatch that is suitable for handsfree barcode scanning, data collection, monitoring, and goods traceability in production lines, retail, and warehouse.[86] This improves the order processing ability and operations efficiency while relieving muscle fatigue encountered in traditional handheld scanning devices. It features a 1.3 GHz Quad Core processor, a 3.35 in. display, an 8MP camera, touch keys, 15 h of battery life, and vibration reminders. The Hyco W27 Ring Scanner is based on a patented scanning module that runs on ultralow power and has been adopted in multiple industries, including Logistics and Express, E-commerce and Warehousing, Manufacturing, Medical Logistics, and Retailing.

Buildots uses 360° cameras attached to the workers' helmets to capture videos images of a construction site.[87] The subsequent digitization of the landscape is used to compare the ongoing progress with the original plans for better scheduling. Its AI-based computer vision features is used to detect and track partially completed tasks, identify visual defects in a timely manner, and calculate expected payments based on the project progress and anticipated hurdles.

Tappy Technologies is a wearable token service provider (WTSP) with global partners in the mobile banking and payment card sectors.[88] Its tokenization technology is used to digitize a variety of information sources (e.g., payment cards, access cards, loyalty cards, transit passes, and event passes) into a convenient near-field communication (NFC)-enabled wearable called Uppu. This small wearable can be attached to keyrings or watchstraps to make contactless payments. Its Digital Wallet companion app monitors the transactions conducted by tapping Uppu and displays the results on a smartphone. To advance this wearable tokenization technology, a partnership between Mastercard, MatchMove, and Tappy was recently announced.[89] On a similar note, the K Ring by Mastercard is the world's first contactless, wearable payment ring for use in retail shops around Europe.[90] To avoid accidental payments, this scratch-resistant zirconia ceramic ring has to be at a specific position and within 4 cm of the reader. The K Ring does not need charging or pairing with any smartphone or smartwatch to make payments. The following are examples of workplace wearables for tracking the activity, behavior and social interactions of workers to boost productivity.

Table 3. Examples of wearable devices to monitor the productivity of workers.

| Company | Sample products | Classification | Components | Productivity applications |
|---|---|---|---|---|
| Epson | Moverio BT-350 | Smart Glasses | Input devices, displays, projection systems, audio-vision and motion tracking sensors, immersive technologies | Field service, vision picking; visualization of instructions; interactive 3D representation; real-time feedback and alerts; remote training and supervision |
| Google | Glass Enterprise Edition 2 | | | |
| Microsoft | HoloLens 2 | | | |
| RealWear | HMT-1 | | | |
| Toshiba | DynaEdge DE-100 | | | |
| Vuzix | M400 | | | |
| Daqri | AR Smart Helmet, Smart Glasses, WorkSense | AR headsets, Hard Hat | AR hardware, 3D depth sensor, 360° navigation camera, Industrial 4D Studio | Visualization of instructions, 3D mapping, training, supervision, real-time alerts |
| Foci | Focus wearable | Clip on device | Motion sensors, neuro-respiration matching algorithms, noise elimination software | Capture, visualize and optimize the cognitive states and current focus levels |
| Humanyze | Sociometric ID Badge | Badges, Tags | Microphones with voice analysis; motion detectors, location tracker; behavioral analytics | Track employee interactions and data flow; predict organizational health and productivity |
| Hyco | Hyco W26 Ring scanner, W562 Smartwatch | Barcode Scanner | 3.5 in. touchscreen display; IP66 protection; pedometer, GPS, WiFi/USB/Bluetooth | Hands-free ring scanner; automated entry and tracking of goods |
| Muse | Muse 2, Muse S, Muse Direct App | Headband | Brainwave feedback (EEG), pulse oximetry, SmartSense Rubber Ear sensors | Measure brain activity; improve mental focus; guided meditation |
| Nymi | Nymi Band, Lynk App, Enterprise Software | Handband | Cryptographic processor, fingerprint scanner, heartbeat and motion sensors, IP66 protection | Workplace security; biometrics; liveness and compliance checks |
| ProGlove | Mark 2 Scanner Series, Insight Web portal | Barcode Scanner | Smart glove, acoustic and haptic feedback; 59 in. scanning range; up to 10 000 scans per battery life | Hands-free barcode scanning for picking, sorting and sequencing; direct parts marking |
| Rufus | Rufus Cuff, ScanGlove, WorkHero Software | Barcode Scanner | Cuff and gloves, 5.5. in. display, IP67 protection, KPIs, productivity analytics | Faster hands-free scanning; inventory and device management; team visibility |
| The Tactigon | Tactigon Skin, Tactigon ONE | Gesture Controller | Gyroscopes, accelerometers, magnetometers, AI algorithms, Arduino SDK, 8 + h battery life | Control gestures and motion of robots, video games, 3D printers, drones, AR/VR computers |
| Thalmic Labs | Myo Gesture Control Armband | Gesture Controller | Sensors to measure electrical activity of muscles, ARM Cortex M4 processor, haptic feedback | Control technology with hand gestures (e.g., robotics from EZ-Robot, Sphero, and Parrot) |
| Thync | Calm and Energy Wearable, Relax Pro | Brain neuro-stimulator | Electrical device for neurostimulation; pads; deep relax and deep sleep modules | Target the autonomous nervous system; boost mental energy; relieve stress and anxiety |
| Xsens | MVN Animate, MVN Analyze, MTw Awinda | Human motion tracker | Inertial motion capture sensors, cameras; fast data sampling devices; ML/AI software | Full-body motion capture; positional tracking of multiple actors; biomechanical modelling |

a) Humanyze uses a wearable badge for monitoring its employees at work, and subsequently employs organizational analytics to understand how teams interact to enhance efficiency and productivity.[91] The Humanyze Badge has sensors to detect whether the wearer is moving or still, in close proximity to other badged users, and talking or not talking. The badge does not record the content or personal activities of users, and the recorded metadata are encrypted for anonymity. Its Organizational Health Platform uses the sensors' data to estimate the levels of engagement, productivity, and adaptability, and provide a macro-view of the degree of collaboration within the organization to identify high-performing trends and areas of improvement. b) The MIT Human Dynamics laboratory used sociometric badges to collect data on people's tone of voice and body language to track social behavior in workplaces and gain insights into effective team building.[92] c) Amazon is developing a wristband device for workers in its warehouse and fulfillment centers.[93] Using ultrasonic devices around the warehouse and within wristbands, the physical location of each worker is tracked in real time.[94] The wristband also vibrates to inform the worker of the direction of the next assignment. d) Tesco, a British grocery chain, uses armbands worn by forklift drivers and warehouse staffers to track goods being transported to its 90 aisles of shelves.[95] These armbands provide a mechanism for tracking the order fulfillment and job completion time. e) Nymi workplace wearables are designed to simultaneously address the issues of workplace productivity, safety, compliance, security, and user privacy.[96] The Nymi Band is a wearable device that provides password-free, biometric worker authentication through sensors for fingerprinting and heartbeat detection. The Nymi Band has an OLED monochrome display (48 × 64 pixels) and a battery life of over three days and. It incorporates secure NFC and Bluetooth Low Energy (BLE 4.2). The Nymi Lynk is a smartphone app that allows full traceability and visualization of privacy data stored on the Nymi Enterprise Server. The Nymi Software Development Kit (SDK) helps developers to build APIs and integrate their existing tools with the Nymi Enterprise Server and Nymi Quality Management System. The Nymi technology is used for a variety of workplace applications including privileged access management, contactless physical access control, transaction authorization, workforce communication, intelligent print solutions, and management of employee health and safety.

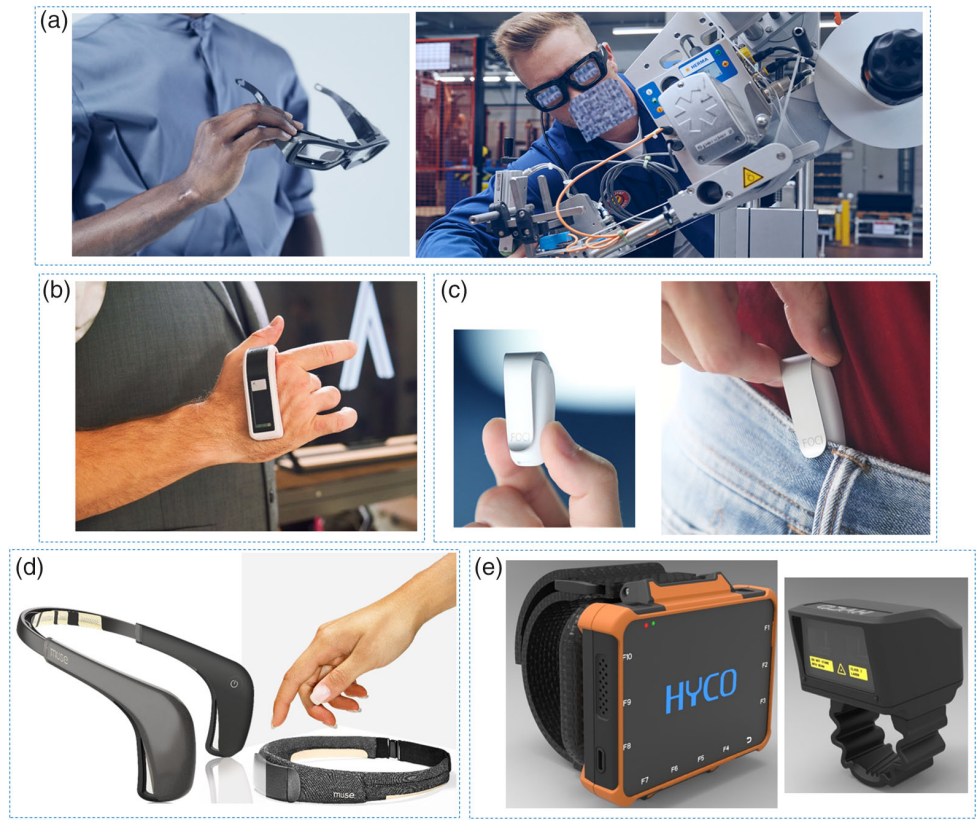

**Figure 3.** Examples of wearables for occupational productivity monitoring. a) ViewPoint Systems VPS 16 smart glasses have stereoscopic cameras that track spatial vision and have live streaming capabilities for mixed reality applications. b) Tactigon Skin uses sensors to detect hand gestures and AI to remotely interact with digital machines. c) The Tinylogic Foci clip-on device monitors the breathing patterns to estimate cognitive state of mind and suggests ways to avoid distraction at work. d) Muse 2 and Muse S headbands measure brain electrical activity and provide real-time feedback on body performance and mental focus. e) Hyco's W562 smartwatch enables hands-free barcode scanning for goods tracking and asset management. a) Reproduced with permission.[231] Copyright 2021, ViewPoint System; b) Reproduced with permisson.[107] Copyright 2021, The Tactigon; c) Reproduced with permission.[121] Copyright 2021, Tinylogic Foci; d) Reproduced with permisson.[114] Copyright 2021, Muse; and e) Reproduced with permission.[86] Copyright 2021, Hyco.

With the work-from-home culture during the COVID-19 pandemic, there has been a surge in surveillance tools for remotely monitoring the workers' activities.[97] Common surveillance methods for office desk workers include keystroke logging, instant messages monitoring, screen monitoring, user action alerts, location tracking, webcam surveillance, and remote control takeover. Some of the rapidly growing providers of employee surveillance tools are Time Doctor, FlexiSPY, Hubstaff, DeskTime, ActivTrak, Teramind, SpyEra, WorkTime, and Kickidler.[97] Recently, Microsoft released a Productivity Score feature for its Office 365 to analyze the amount of time a user spends on its software (e.g., MS Outlook and Team).[98] Microsoft has stated that this opt-in feature will provide insights on how people and teams use its technology and infrastructure. However, privacy advocates have raised concerns about this controversial feature as it appears to be a remote surveillance tool for monitoring worker activity at an individual level.[98]

### 4.2. Augmented Reality and Virtual Reality

Depending on the specific AR/VR application, head-mounted displays come in different configurations (i.e., monocular or binocular, transparent or nontransparent displays). Common features of industrial-grade AR/VR smart glasses include accurate voice recognition, hands-free navigation, gesture recognition, large and flexible display, lightweight construction, comfortable fit, rugged design, and water- and dust-resistance. Typical parameters for comparing industrial-grade AR/VR glasses are the pixel resolution (e.g., 1280 × 720 pixels per eye), field of view (10–100°), refresh rate (90 Hz–144 Hz), tracking and control (eye and hand), battery life (2–8 h), and pricing ($50–$3500). Examples of AR/VR smart glasses include the Google Glass Enterprise Edition 2, Microsoft HoloLens 2, Vuzix M400, RealWear HMT-1, Epson Moverio BT-350, and Toshiba DynaEdge DE-100. The following are application examples of where AR/VR smart glasses have been used to boost productivity.

The Google Glass has been used by employees at Boeing and Tesla to receive complex instructions in their assembly plants and accelerate their production using voice, graphics, and gesture-controlled interactions.[99,100] Boeing employees install complex electrical wirings for airplanes using hands-free, interactive 3D representations of wiring models shown by Google Glass. Such installation would otherwise be tedious and error-prone using 2D drawings. Google Glass has been used for a variety of industrial applications, such as to monitor the plant processes or path feed rates, real-time calibrations and fault checks, view and send alarms/messages, and share/capture/store live streaming videos or still images. The Google Glass Enterprise Edition 2 has a Qualcomm Snapdragon XR1 (up to 1.7 GHz), Android Oreo, 8MP 80° field-of-view camera, three beam-forming microphones, 3 Gb LPDDR4, 820 mAh battery, 6-axis accelerometer and gyroscope, USB Type-C port, and on-head detection sensors.

DHL Supply Chain has used AR smart glasses (i.e., Google Glass, Vuzix M100 and M300 and UbiMax software) for its "Vision Picking Program" initiative.[101,102] The AR smart glasses has been used by DHL for hands-free barcode scanning, voice recognition, pickup location display, single and multiunit picking, best pick path recommendation, visual guidance through picking and kitting tasks, and heads-up display with an intuitive display. All of this has positively affected the DHL supply chain including warehouse planning, freight loading, transportation, completeness checks, dynamic traffic support, and parcel loading or drop-off.[103]

Pick-by-Vision solutions has been developed by Picavi. Its smart glasses (i.e., Glass Enterprise Edition, and Vuzix M300) and Picavi Power Control has been used for faster and more reliable material flow by providing guided object picking in warehouses to ensure efficient intralogistics and minimum error rate.[104] The Picavi ecosystem has a unique feature of push notifications with prioritization of messages which allows bidirectional, contactless, and individual communication between the floor worker/picker and the staff in the control room. This facilitates to report any problems, reprioritize operations, and correct any actions without much delay.

The Voxware AR brings a voice management solution to boost agility, efficiency and productivity in distribution centers.[105] Voxware combines the AR vision technologies of smart glasses with voice and scanning features to automate the workflow in distribution centers, including the picking, packing, cycle counting, returns and receiving, put-away of inventory, and real-time supervisory monitoring. The VoxPilot Analytics platform aggregates Voxware data to provide real-time visibility and insights into future events. The VoxPilot platform has been used to optimize inventory positions, increase traceability of packages, streamline their predictive analytics, and eventually produce a 10–15% increase in productivity.

Daqri has developed the Daqri Smart Glasses that work with a suite of AR productivity tools within the Daqri WorkSense platform. The smart glasses consist of multiple cameras (depth sensor camera, color camera, and AR tracking camera), and is used by the wearer to upload 3D models, navigate around them, and zoom in/out at the wearer's own pace with precise positional tracking and reliable interactions with the surroundings. The Daqri Smart Helmet is an Android-based hard helmet with AR-driven smart assistance features used by industrial workers for the visualization of instructions, training and supervision, video capture, 3D mapping, and real-time alerts.

The Raptor Smart glasses by Everysight employs AR to make their users aware of their outdoor surroundings, display routes for real-time navigation, and record videos with customizable audio inputs.[106] The smart glasses use a combination of five different sensors, a Qualcomm SnapDragon 410E processor, 16 or 32 Gb memory, and proprietary BEAM display technology to project an added AR information layer to the wearer. There are options to connect with other Bluetooth or BLE-enabled sensors such as heart rate monitor, power meter, and speed sensors. The smart glass weighs ≈98 g or 3.5 ounces, and a full charge can give around 8 h of ride time. Additional favorable features of the Raptor Smart glasses for outdoor work activities are their aerodynamic design, smartphone connectivity, light and dark tint visor, protection against high mass/velocity impacts, and total privacy.

### 4.3. Gesture and Motion Control

Gesture and motion control wearables offer hands-free interactions with the digital world to enable AI-driven complex 3D functionalities such as remote control of AR/VR machines, mechanical or surgical robots, mobile payments, biometrics, assembly lines, retail business, sign language interpretation, medical-assistive living, and 3D animation. Robotic actions are enabled are detecting various hand or finger gestures such as arm stiffening, fist clenching, hand rotation, wrist flexion and extension. Typical benchmarks used to compare different devices in this category are price, compatibility (Linux, Mac, Windows), wireless connectivity (2.4 GHz), degrees of freedom (6–10), hands freedom, left- and right-hand operations, gesture recognition, voice control, battery life, programmable function keys, SDK (Arduino, Python), and pairing with multiple devices. Examples of wearables of gesture and motion control are as follows.

The Tactigon Skin device is essentially a 3D wearable mouse that recognizes voice commands and the gestures of hand and fingers. It is used to remotely control the functions of mechanical robots, drones, rovers, AR/VR computers, 3D CAD design and presentation workflow, and 3D printers.[107] It incorporates sensors (three-axis accelerometer, three-axis gyroscope, and three-axis magnetometer) that detect linear and angular hand movements, and uses artificial intelligence (AI) to convert them into actions for digital machines. The Tactigon ONE is an Arduino-compatible microcontroller board that can be further programmed by the user to implement new gesture and motion control features. There are over 48 complex gestures and hand movements that can be recognized by the Tactigon Skin, including gestures to grab close/open, stretch in/out, forward/backward, up/down, roll right/left. Pitch up/down etc.

The Myo Gesture Control Armband by Thalmic Labs is designed to control digital technologies using hand gestures.[108] The Myo armband has sensors that detect hand movements and the electrical activity of hand muscles. The incorporated sensors to sense motion in any direction are a three-axis gyroscope, three-axis accelerometer, and three-axis magnetometer. There are eight medical-grade electromyography (EMG) sensors. The detected hand gestures are used to generate appropriate control signals for digital tools and gadgets such as prosthetic forearms, AR drones, and VR gaming consoles. Some specific gestures enabled by the Myo armband are double tap (to enable or disable mouse movements), wave in or wave out (to zoom in or zoom out of the object), spread fingers (go to the next object), and fist (rotate an object by arm movement, and enable or disable rotation mode).

Xsens has a number of products related to full-body human motion capture systems.[109] The Xsens MVN Animate and Xsens MTw Awinda incorporate wireless inertial sensor modules to track the positions of multiple actors in real time with time-synchronized data sampling (of within 10 μm). Xsens Sensor Fusion algorithms work with the Attitude and Heading Reference System and Inertial Navigation System to determine the 3D orientation and relative position after compensating for gravity and Earth's magnetic field. The Xsens MVN Analyze software platform integrates the incoming data streams from motion capture systems for easy visualization in real-time. The Xsens products are used to create 3D animations and visual effects (VFX) effects for movies, gaming, TV broadcasting, and live entertainment. They are also applied in performance science, injury prevention, biomechanical modeling, and workplace ergonomics.

### 4.4. Mental Acuity, Brain-Wave Sensing, and Occupational Stress Management

The work-related risk factors for poor mental acuity are occupational stress, work shifts, irregular working hours, bright lights, odors, and noise.[110] These risk factors can also stem from job insecurity, competition, low psychosocial support, low decision-making latitude, demanding job profile, and work-life imbalance.[111] Consequently, productivity can be compromised with risks of injuries, errors, absenteeism, and irregular attendance.[112,113] The following are examples of brain-wave sensing headbands for improving workers' mental acuity.

The Muse 2 headband is a compact electroencephalography (EEG) system that assists in the art of mental focus.[114] The multisensor Muse headband detects and measures brain electrical activity and provides real-time feedback on body performance (i.e., movement, heart rate, sleep activity, and mental activity). It has a weight of 41 g, LED indicator, micro-USB for charging, gold/silver sensors, comfort-fit fabric (90% Raylon, 5% Nylon, and 5% Spandex), and a rechargeable lithium-ion battery (10 h battery life). The Muse Direct App uses ML to decompose the raw brain-wave data into understandable components such as oscillations, transients, noise, and event-driven responses. The Muse app makes it possible to listen to streaming brain activity in the form of audio clips or immersive soundscapes.

Melon has designed a smart headband with a mobile app to track the mental focus during any activity with respect to one's emotions, behavior, and surroundings.[115] Melon's headband consists of a wireless brain-sensing device that measures EEG signals and correlates them with brain activity. Melon's headband incorporates a EEG monitoring chip from NeuroSky that has expertise in EEG signal amplification, electrical noise reduction, and filtering techniques.[116] The Melon mobile app uses algorithms to quantify focus and provides personalized feedback on ways to improve focus during activities at work, study or play. Melon's headband technology was acquired by Daqri that plans to integrate the EEG monitoring technology into their smart helmets built with AR capabilities.

The wireless EEG headset by IMEC is capable of detecting emotions and attention at real-time.[117] The headset comes with pre-fitted dry electrodes to record the brain's electrical activity, whereas its built-in software helps to visualize brain-wave signals. This clinical-grade IMEC headset is used to improve cognitive abilities through remote sensory stimulation and VR-based treatment of cognitive conditions (e.g., autism and attention-deficit/hyperactivity disorder (ADHD)).

SmartCap Technologies has developed a LifeBand device to measure mental fatigue and preventing microsleep at work.[118] The LifeBand device is attached to a headwear (e.g., helmet, cap, or beanie) to record brain waves and infer the alertness or mental fatigue levels. Its fatigue speedometer and brain-wave monitoring band are suitable for road transport drivers, pilots, train drivers, miners, and machine operators.

In addition to brain-wave sensing, alternative measurements are being sought to monitor and manage occupational stress, for example, by sensing electrodermal activity and respiratory patterns as discussed here.[119] a) Pip has developed a finger-wearable electrodermal activity monitoring device for stress monitoring and management.[120] The skin pores at the fingertips are sensitive to stress, and the Pip device captures this stress-related variation by monitoring the electrodermal activity at our fingertips. Stressful events can trigger several physiological changes, including the sweat gland activity that is picked up by the Pip device. My Pip and other companion apps allow the user to visualize the changing stress levels, suggest relaxation sessions to externalize the body stress, compare your progress over time, and store personal data on a cloud platform that complies with the Health Insurance Portability and Accountability Act (HIPAA). b) The Focus Wearable clip-on device by TinyLogics Foci helps to better visualize the current state of mind by monitoring respiration.[121] Foci technology uses motion sensors to detect tiny movements in breathing and employs ML algorithms to correlate the breathing patterns with cognitive states of mind. The Foci app uses advanced emotion recognition and emotion elicitation models to display the state of mind (categorized as relaxed, fatigue, or stressed) and suggests ways to tune out distractions, terminate procrastination, sustain longer focus, and acquire a balanced psychological flow.

The next frontier in occupational stress management lies in building personalized models with appropriate baselines of clinical disorders and pre-existing conditions. This is because physiological stress response (both type and magnitude) strongly varies across individuals as suggested by a large stress detection study involving 1002 individuals and conducted over five consecutive days.[122] In addition to occupational stress, sleep is another important determinant of cognitive failures and reduced mental acuity in the workplace.[123,124] While polysomnography (PSG) is more accurate than actigraphy, most consumer wearables (i.e., smartwatches and smart rings) are sufficient to monitor daily sleep attributes.[125–127] A recent study found that the performance of the Samsung Gear Sport watch and Oura smart ring were comparable with that of a medically approved actigraphy device in tracking total sleep time, sleep efficiency, and wake after sleep onset.[128] Philips Respironics has also developed a suite of motion biosensors to track sleep–wake cycles and circadian rhythms using the Philips Health band, ActiWatch Spectrum Pro, and ActiWatch Spectrum Plus.[129] Its Actiware software provides a unified platform for centralized management of data from multiple actigraphy devices.

## 5. Wearables for Occupational Health Monitoring

In this section, the following categories of wearable devices for monitoring occupational health are discussed: i) work-related musculoskeletal disorders, ii) functional movement disorders, iii) occupational lung diseases and continuous respiratory monitoring, iv) occupational cardiovascular diseases, and v) occupational sun protection, thermal comfort and continuous glucose monitoring (CGM). **Table 4** shows the commercial devices for monitoring worker health the companies that produce them, their classification, components, and applications. **Table 5** similarly shows the devices for monitoring the physical activity of workers. **Figure 4** and **5** shows examples of wearable technology products for monitoring occupational health and physical activity, respectively.

**Table 4.** Examples of wearable devices to monitor the health of workers.

| Company | Sample products | Classification | Components | Health applications |
|---|---|---|---|---|
| Arion | Smart insoles, Footpods, coach | Footwear | Pressure sensors in insoles, footpods, accelerometer, gyroscope, dashboard | Mobile analysis of gait, body movement and running metrics; real-time feedback |
| Cyrcadia Health | iTBra, Cyrcadia Breast Monitor | E-textiles, Patch | Two sets of eight digital temperature sensors, Bluetooth-enabled data recording device | Detect early breast tissue abnormalities; monitor thermo-circadian clock rhythms |
| Embr Labs | Wave Bracelet | Thermostat | Thermal skin sensors and regulators | Warming or cooling the inner region of the wrist |
| Everysight | Raptor, Controller, Rx Adapter | Smartglasses | AR glasses, BEAM display projector, motion sensors, microphones | Visualize and be aware of outdoor environments; route display; cycling computer |
| Halo Neuro | Halo Sport 2 | Headset | Primer band, electro-neurostimulator | Hyperplasticity of brain motor neurons |
| K-invent | K-invent Grip, Sens, Controller, Kforce App | Force and Flexibility monitor | Goniometer, force plate sensors, dynamometer, accelerometer, gyroscope | Monitor flexibility of body limb joints, isometric strength, grip strength, range of motion |
| L'Oreal | UV Sense, My Skin Track | Patch | Sensors for UV, particulates, temperature | Measures UV exposure, pollution, heat, humidity |
| Lumo | Lumo Run, Lumo Lift | Clip-on device | 9-axis IMU, accelerometer, gyroscope, barometer, magnetometer, vibration motor | Correct the running form and posture through vibrational feedback and coaching |
| MagnetRX | Magnetic Therapy Bracelet, Cuff | Wellness Jewelry | 40 powerful rare-earth neodymium magnets, titanium with ion plating | Magnetic therapy, improve circulation, reduce toxins, relieve inflammation and pain |
| Marakym | Clavicle Brace | Adjustable braces | Upper body straightener, posture feedback | Posture corrector; reduced shoulder and back pain |
| Omron | HeartGuide, KardiaMobile | Wristwatch, Finger pads | Inflatable cuff, transflective memory-in-pixel LCD, EKG sensors | Correlate blood pressure to daily activities and sleep patterns; detect EKG and heart health |
| Owlet Care | Smart Sock, Cam, App | E-textile (socks) | Motion sensors, pulse oximeter, base station; 16 h battery life | Tracks baby's heart rate, blood oxygen levels, and sleep trends; For home use or daycare |
| Pip | The Pip, My Pip App | Wearable device (finger) | Sensors to measure electrodermal activity and stress, biofeedback mechanism | Stress monitoring and management; overall stress-related health improvements |
| QSun | QSun Wearable, App | Clip-on device | Electronic UV sensor, temperature sensor | Tracks UV exposure levels, vitamin D production |
| Qardio | QardioArm, QardioBase | Wearable device (arm) | Inflation and pressure release valves, blood pressure and heart rate sensors | Measure systolic and diastolic blood pressure, heart rate; Irregular heart beat detection |
| Spire Health | Spire Health Tag | Clip-on device | Respiration sensor for thoracic excursion, three-axis accelerometer, PPG pulse-rate sensor | Measure respiratory patterns, stress, pulse rate, activity, sleep duration, sleep quality |
| Vital Connect | Vital Patch, Solution 2.0 | Adhesive Patch | Three-axis MEMS accelerometer, ECG electrodes | Continuous patient monitoring with prediction |
| Wearable X | Nadi X pants, Pulse | Clip-on device | Integrated sensors, 370 mAh battery | Monitor yoga posture, vibrational guidance |
| Withings | Withings Sleep, Health Mate App | Sleep Tracking Mat | Integrated sensors in the mat, Bluetooth 4.0 | Track sleep cycles (deep, light, REM), sleep duration, heart rate, snoring duration |

Table 5. Examples of wearable devices to monitor the physical activity of workers.

| Company | Sample products | Classification | Components | Physical activity applications |
|---|---|---|---|---|
| Apple | Apple Watch Series 6 | Smartwatch, Activity Tracker, Band | Three-axis accelerometer, gyroscope, barometer; Optical heart rate sensor, capacitive sensor, photodetector, microphone, connectivity (Bluetooth 5.0, NFC, LTE, WiFi b/g/n, A-GPS, GLONASS), battery (205 to 470 mAh), operating system (watchOS, Wear OS, Tizen, AsteroidOS) | Tracking activity, running, fitness, performance, heart rate, sleep, stress, breathing, water intake, falls, and altitude acclimation |
| FitBit | Versa 2, Charge 4 | | | |
| Fossil | Gen 4, Gen 5, Hybrid HR | | | |
| Garmin | Vivoactive, Vivofit, Vivosport | | | |
| Microsoft | Microsoft Band 2 | | | |
| Mobvoi | TicWatch Pro, Sport, Express | | | |
| Philips | ActiWatch Spectrum | | | |
| Polar | Unite, Grit X, A370, M430 | | | |
| Samsung | Galaxy Watch, Active2 | | | |
| Atlas | Multi-Trainer | Performance Tracker, Compression Sleeve, Band | Sensors for movement and activity; pressure sensor, heart rate sensor, sleep monitor; algorithms for movement patterns and exercise routines | Detect movements through space (handstand walks, box jumps, kettlebell swings); track gym activities (moves, reps, sets); monitor HRV, sleep and muscle fatigue |
| Casio | G-Shock DW, G-MS, MT-G | | | |
| Garmin | Forerunner 945, Fenix 6 | | | |
| Push | Nexus | | | |
| Wahoo | Tickr, TickrX, TickrFIT | | | |
| Whoop | Racer, Power, Impact | | | |
| ActivInsights | GENEActiv Original, Sleep, Action, Wireless | Wrist band | Three-axis accelerometer; sensors for physical activity, sleep and wake cycles, ambient light, temperature | Lifestyle reporting; tracks physical activity, sleep, behavioral changes and performance |
| Catapult | Catapult Vector, ClearSky, PlayerTek | E-textile | Accelerometer, gyroscope, magnetometer, heart rate monitor; 6 h battery life | Sports training and insights; monitor performance, risk, and return to play |
| Hexoskin | Hexoskin Smart Shirt, Astroskin Platform | E-textiles | Sensors for ECG and heart rate, QRS events, breathing rate, activity intensity, sleep; 30 + h of battery life | Track data on cardiac, respiratory, activity, sleep parameters |
| Motiv | Motiv ring | Smart Ring | Flexible circuit board, curved battery, heart rate sensor; titanium shell; 3 day battery life | Track activity (type, duration, intensity), heart rate, calories burnt, sleep duration |
| Oura | Oura ring | Smart Ring | Flexible circuit board, infrared LED, temperature sensor, accelerometer, gyroscope; titanium body Negative Temperature Coefficient | Monitor body pulse, movement, temperature, sleep quality |
| Sensoria | T-shirt, sports bra, socks | E-textiles, Apparels | 9-axis MEMS sensors, heart rate monitor, textile electrodes, ultralow power SoC | Measure heart rate with upper garments; Measure cadence, impact forces, foot landing |
| Under Armor | UA band, UA heart rate, UA record | Wrist band | Sensors to monitor heart rate, movement, activity | Track sleep, step count, heart rate, workout |

## 5.1. Work-Related Musculoskeletal Disorders

Musculoskeletal disorders negatively affect the natural body movement and musculoskeletal system (i.e., muscles, ligaments, tendons, nerves, discs, and blood vessels).[130,131] The risk factors for work-related musculoskeletal disorders (WMSDs) can be ergonomic (i.e., due to manual lifting, forceful exertions, repetitive tasks, or awkward posture) or individualistic in nature (i.e., poor habits, lack of rest and recovery, and unhealthy work practices).[132] Exposure to these risk factors causes fatigue, imbalance, discomfort and pain which can worsen to WMSDs and loss of function. Upper body disorders (including neck and back pain) are common in office jobs where poor sitting postures, a sedentary work style, low social contact, and high stress are prevalent.[133,134] Lower back pain is triggered by ergonomic stressors generally encountered by workers in farming, manufacturing, construction, packaging, transportation, and service desk jobs.[135–137] Carpal tunnel syndrome results from repetitive tasks and specific wrist positions, leading to pressure on the wrist's median nerve and associated pain, numbness, and weakness in the hand and fingers.

Wearables for WMSDs assess the natural gait, posture, and physical strength in an objective and nonintrusive manner. Two examples are as follows: a) The McRoberts DynaPort MiniMod system consists of the MoveTest and MoveMonitor for monitoring exercise capacity, physical activity, and gait parameters.[138] McRoberts products are used with standardized tests such as the Six Minute Walk Test (i.e., total distance walked in 6 min, walking speed, and step length and frequency), Sit-to-Stand Test (i.e., time to rise 5 times from seated position on a chair), GAIT Test (i.e., gait parameters during a straight trajectory), Sway Test (i.e., displacement in the center of pressure), and Timed Up and Go Test (i.e., time for sequentially rising from a chair, gait initiation, walking, and sitting down on the chair). b) K-invent wearable products are designed to monitor the progress of rehabilitation interventions to regain or improve the lost body functions.[139] Examples of K-invent sensors are force plates that track posture and lower limb strength, grip dynamometer to measure grip strength, and goniometers that measure the range of motions by body joints. The companion Kforce app is used to visualize, customize, and manage the rehabilitation regimen.

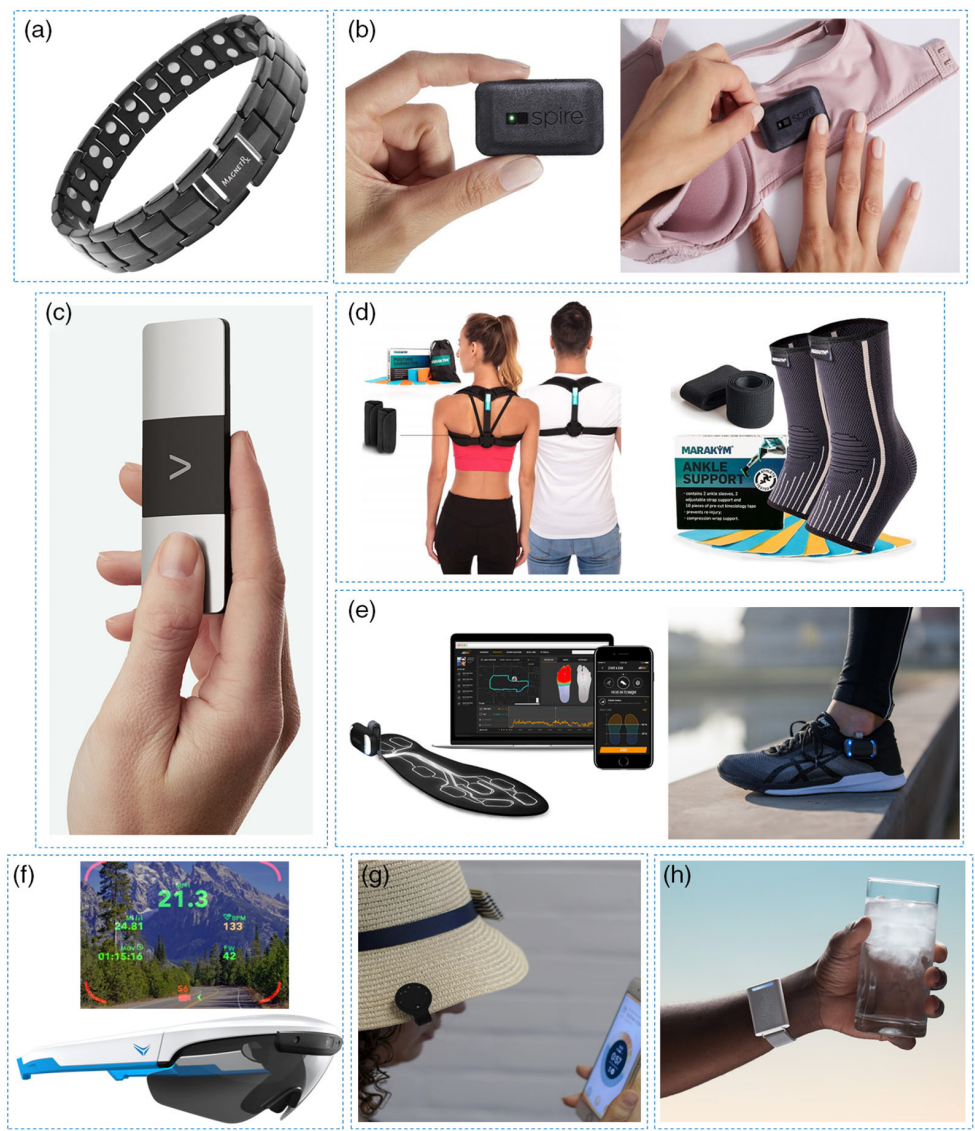

**Figure 4.** Examples of wearables for occupational health monitoring. a) The MagnetRX Ultra Strength Magnetic Therapy Bracelet uses magnetic field stimulation to reduce pain and inflammation. b) The Spire Health Tag tracks the breathing, sleep, and stress patterns with advice on stress reactivity. c) Omron's KardiaMobile 6L EKG detects heart rhythm, atrial fibrillation, bradycardia, and tachycardia. d) The Marakym Posture Corrector ensures stability and alignment, while the Marakym Ankle Support uses compression to prevent injury or facilitate faster recovery from injury. e) Arion Smart Insoles and Footpods monitor the cadence, posture, and balance with real-time feedback on corrective measures. f) Everysight Raptor Smart Glasses assist with navigation displays and outdoor awareness. g) The QSun UV Exposure Tracker monitors outdoor sun exposure and vitamin D levels with personalized recommendations. h) The Embr Labs Wave Bracelet generates warm or cooling waveforms on the wrist to alleviate thermal discomfort. a) Reproduced with permission.[150] Copyright 2021, MagnetRX; b) Reproduced with permission.[177] Copyright 2021, Spire Health; c) Reproduced with permission.[186] Copyright 2021, Omron; d) Reproduced with permission.[232] Copyright 2021, Marakym; e) Reproduced with permission.[67] Copyright 2021, Arion; f) Reproduced with permission.[106] Copyright 2021, Everysight; g) Reproduced with permission.[196] Copyright 2021, QSun; and h) Reproduced with permission.[197] Copyright 2021, Embr Labs.

Activity trackers (e.g., smartwatches and smart rings) represent an immensely popular class of consumer wearables that have overlapping relevance to occupational health and fitness (Table 5).[140] This is because physical activity is beneficial both at home and the workplace as it enhances musculoskeletal strength, speed, stamina, accuracy, balance, sleep quality, and cardiovascular and respiratory endurance. The Physical Activity Guidelines for Americans issued by the USA. Department of Health and Human Services recommends that adults obtain at least 150 min per week of moderate-intensity aerobic activity and at least 2 days of muscle-strengthening activity.[141] A recent harmonized meta-analysis of more than 44 000 men and women using activity trackers concluded that 30–40 min of moderate-to-vigorous intensity physical activity every day reduces the mortality risks associated with sedentary lifestyle.[142] The top sellers of smartwatches are Apple, Fitbit, Huawei, Xiaomi, Fossil, Garmin, Microsoft, Mobvoi, Polar, and Samsung. The market for designer smartwatches is emerging with new products from Kate Spade, Louis Vuitton, Michael Kors, Skagen, and Tag Heuer. A subclass of wristwear and compression sleeves is tailored to track performance and muscle fatigue during specific physical activities, and these products are sold by Atlas, Casio, Garmin, Push, Wahoo and Whoop. Typical parameters for comparing smartwatches are price, weight, size, screen size and resolution, display type, internal storage, battery size and longevity, connectivity, CPU, dedicated GPS, operating system, waterproof ratting, and in-built sensors (e.g., accelerometer, gyroscope, camera, heart rate sensor, light sensor, pedometer, barometer, pulse oximeter, speaker, and microphone). Recent studies have used smartwatches to track physiological signs (e.g., body temperature, respiration rate, heart rate, heart rate variability, sleep, step count, etc.) for the early detection of COVID-19 symptoms and to predict the likelihood of infection.[143–146] Amazon has just announced its foray into health and fitness wearables with the Halo band for tracking users' activity levels, sleep, body fat percentage, and tone of voice.[147] Using body photos and sound bytes from a microphone, the Halo companion app generates a 3D rendering of the user's body for a holistic analysis of body fat and tone.[148]

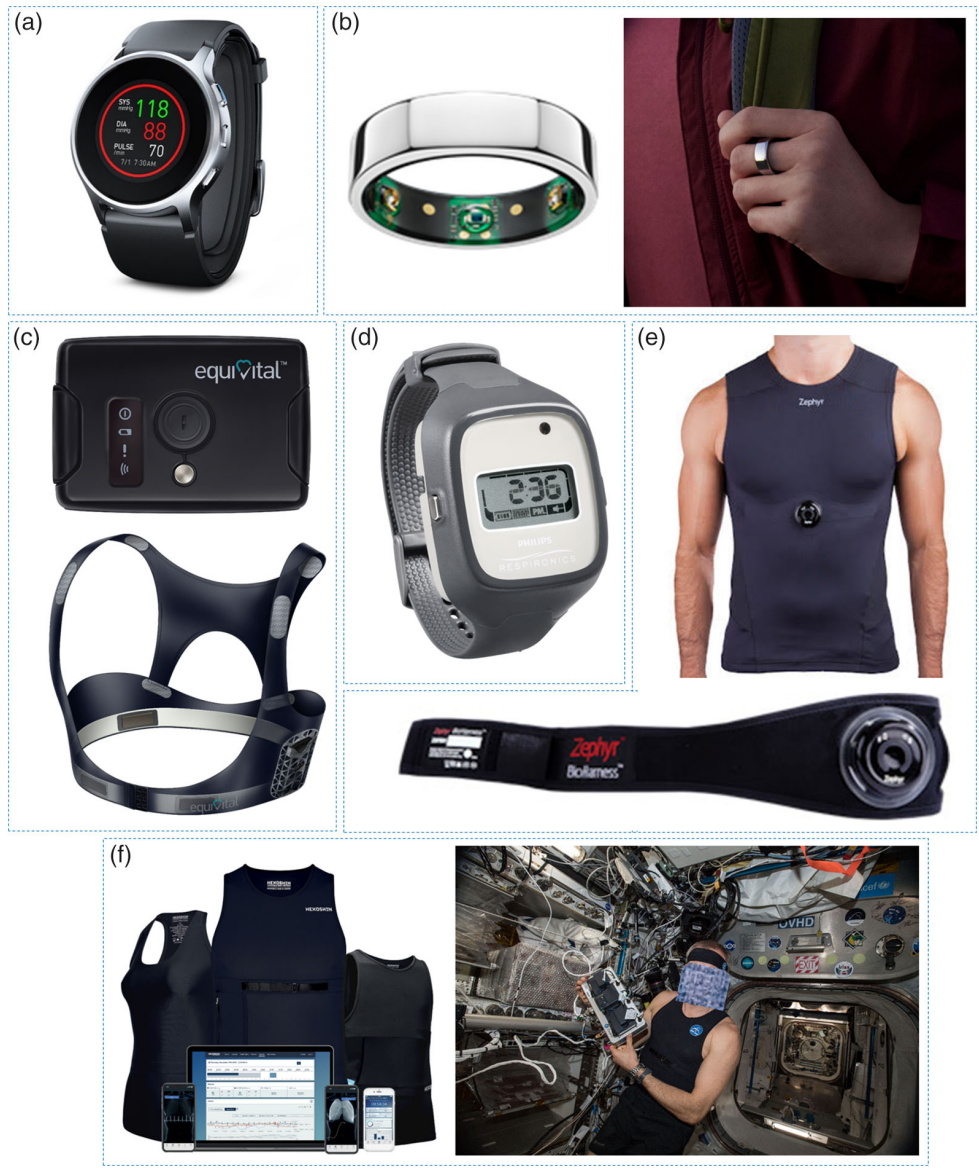

**Figure 5.** Examples of wearables for physical activity monitoring. a) Omron's HeartGuide BP8000-M tracks the user's aerobic steps, calories burned, and blood pressure. b) The Oura Ring monitors the heart rate, body temperature, activity, calories burned, respiratory rate, and sleep. c) Equivital's LifeMonitor is a mobile human monitoring system that continuously measures the user's galvanic skin response, respiratory rate, heart rate, ECG data, and activity. d) The Philips Actiwatch Spectrum Plus measures activity and cycles of being awake or asleep. e) The Zephyr BioHarness is a physiological monitoring device with telemetry for tracking the activity, body orientation, ECG data, heart rate, and breathing rate. f) Hexoskin Smart Garments monitor the user's cadence, activity, acceleration, heart rate, and breathing rate. a) Reproduced with permission.[186] Copyright 2021, Omron; b) Reproduced with permission.[229] Copyright 2021, Oura; c) Reproduced with permission.[83] Copyright 2021, Equivital; d) Reproduced with permission.[129] Copyright 2021, Philips; e) Reproduced with permission.[81] Copyright 2021, Zephyr; and f ) Reproduced with permission.[233] Copyright 2021, Hexoskin.

Wearables with magnetic field stimulation are used to alleviate work-related musculoskeletal injuries and disorders. Earth Therapy offers copper magnetic bracelets to relieve joint pain, arthritis, carpal tunnel syndrome, migraines, and fatigue.[149] Its bracelets contain six rare-earth magnets delivering over 15 000 Gauss, and is suggested to fight pain and inflammation by removing toxins and improving blood circulation. Copper magnetic bracelets for pain relief are sold by Rainso, Feraco, Reevaria, Juccini, Coppervast, Viterou, and Smarter Lifestyle. Copper knee braces are sold by YD-Copper Health, Thx4Copper and UProtective for pain relief from arthritis, joint pain, strains, tendonitis, and inflammation. The magnetic therapy products by MagnetRX are designed to improve blood circulation and reduce inflammation, migraines, arm pain, and toxins.[150] Its bracelets are built from medical-grade titanium and 40 powerful rare-earth neodymium magnets (3000 Gausses per magnet).

## 5.2. Functional Movement Disorders

Poor motor functions can manifest as hyperkinetic (i.e., excessive, repetitive or involuntary movement) or hypokinetic movement disorders (i.e., reduced or lack of movement).[151] They are mostly caused by neurologic conditions (i.e., degeneration of the brain, brain stem, spinal cord or nervous system).[152] Examples of functional movement disorders include ataxia (i.e., loss of muscle coordination), dystonia (i.e., involuntary muscle contractions), spasticity, tremors and essential tremor, restless legs syndrome, multiple system atrophy, Tourette syndrome, Huntington disease, and Parkinson's disease.[153] The risk factors for functional movement disorders include aging, infection, drug reactions, toxins, previous surgery, stressful events, secondhand tobacco smoke, solvents, metals, and pesticides.

In the workplace, the early signs and symptoms of movement disorders are difficult to discern as the underlying conditions evolve and generally worsen over time.[154,155] The affected person may suffer from postural imbalance, lack of coordination, an unsteady gait, twitching, muscle spasms, and tremors.[156] There may be negative impacts on speech, vision, writing, and eye movement.[157] The signs and symptoms of movement disorders tend to dampen the alertness, energy level, productivity, and social morale of the affected worker.[158] The following are examples of wearables for assessing movement disorders.

Electronic walkways from GAITRite measure the spatiotemporal parameters of gait, compare gait imbalances, document slips and falls, and evaluate the effectiveness of different interventions.[159] As the patient ambulates over the GAITRite walkway, the software captures the relative nature of every footfall. The recorded data are used to establish the baseline functions and to compare with subjective findings for appropriate decisions.

Great Lakes NeuroTechnologies has a range of products for objectively measuring movement disorders, particularly those related to Parkinson's disease.[160] Its products (i.e., KinesiaU and Kinesia 360°) utilize a Kinesia sensor worn on the finger, the Kinesia One App loaded on a smartwatch or iPad, and the Kinesia ProView data visualization toolkit for task-based or continuous motor assessment. The Kinesia sensor consists of a triaxial accelerometer, triaxial gyroscope, Bluetooth radio, and wireless recharging unit. The Kinesia One App is used to report the motor scores for multiple tasks/parameters such as arms resting, arms extended, finger-to-nose, finger tapping, toe tapping, leg lifts, walking and turning, hand grasping, etc. The measured outcomes are resting tremor, postural tremor, kinetic tremor, bradykinesia (speed, amplitude, and rhythm), leg agility, gait, freezing of gait, and/or dyskinesia.

The Personal KinetiGraph (PKG) System by Global Kinetics assess the symptoms of Parkinson's disease (e.g., bradykinesia, dyskinesia, and tremor compared with controls) in a continuous, objective, and ambulatory manner.[161] In addition, the PKG system provides an assessment of daytime somnolence, immobility, and propensity of impulsive behavior. The PKG system consists of a wrist-worn PKG Watch that records movement and proprietary data-driven algorithms that report the findings based on routine daily activities. The PKG Watch is intended to be worn for a 7-day period, after which the processed data (such as scores representative of symptoms, motor fluctuations, sustained immobility, daytime sleep, somnolence, and tremor) are reported to the wearer.

With changing work styles, wearables for posture awareness and real-time feedback to correct slouching or slumping have gained considerable attention. Upright posture helps the oxygen intake of lungs, reduces stress, and boosts the overall body appearance, confidence, strength, vitality, and work productivity.[158] On the contrary, poor posture can lead to physical fatigue, trips and falls, varicose veins, cardiovascular disease, and digestive issues.[162] Posture correctors in the form of upper body braces and straps are available from companies such as BraceAbility, Gearari, Marakym, Truweo, Vokka, and Xnature. Strapless posture correctors from Upright Posture Trainer are designed as patches embedded with multiple sensors to naturally help achieve good posture habits.[163] Its Upright Go 2 device is placed on the back using a hypoallergenic adhesive to track the body posture throughout the day and train the user to correct slouching through gentle vibrations.

## 5.3. Occupational Lung Diseases and Continuous Respiratory Monitoring

The risk factors for occupational lung disease are work-related exposure to gases, dust, fumes, chemicals, sudden temperature changes, and psychosocial stress.[164] Healthcare workers have increased risks of exposure to respiratory pathogens and infectious agents. Chronic obstructive pulmonary disease (COPD) results from the inhalation of smoke and airborne particles commonly found in coal mining, metal smelting, construction, transportation, farming and the manufacturing of chemicals, rubber, leather, plastics, textiles, and food products.[165–172]

Pulmonary function tests are commonly used to assess lung functioning by measuring parameters such as tidal capacity, tidal volume, residual volume, total lung capacity, expiratory forced vital capacity (FVC), forced expiratory volume in one second (FEV1), and forced expiratory flow.[173] A spirometry test measures the airflow breathed in and out into a mouthpiece using the FVC and FEV1. A lower FEV1, FVC, or FEV1/FVC ratio may suggest restrictive or obstructive airways.[174] A plethysmography test measures lung volume by estimating changes in pressure while sitting inside a clear, airtight chamber. Wearable inductive belts, strain sensors, and optoelectronic devices measure changes in the circumference of the ribcage and torso during respiration to estimate the respiration rate and volume.[175,176] The following are examples of wearables for continuous respiratory monitoring.

Spire Health has developed a wearable breathing tracker for chronic respiratory diseases.[177] The Spire Health Tag is used to monitor the breathing patterns, sleep patterns and stress levels, and subsequently provides real-time advice on stress reactivity (e.g., being calm and deep breathing). Its respiratory sensor measures the expansion and contraction of the thoracic cavity, while its photoplethysmography (PPG) pulse rate sensor measures the resting heart rate and calories burned. The Spire Health Tags last for up to 12 months and are designed to be washed and dried. A companion Spire Health app displays the long-term respiration profile (e.g., respiration rate, time for inhalation/exhalation, variability, and full waveform), pulse rate, and activity signature of the wearer. Any deviations from the clinical baseline are automatically detected and shared with the care team many days before symptoms are reported by the patient.

IMEC has developed a wearable patch for continuous respiratory monitoring, even during sleep or ambulatory movement.[117] The patch incorporates its system-on-chip (SoC) readout circuits to record sensor/signal level multimodal signals relevant to respiratory health, such as electromyograms, mechanomyograms, electrocardiograms, bioimpedances, and sound. Using AI algorithms, the inferred noninvasive digital biomarkers are respiratory rate, relative volume, breathing efficiency, oxygen saturation, and respiratory sounds. IMEC has also developed medical algorithms to handle noisy data and provide actionable insights.

In addition, portable, fingertip pulse oximeters are available to measure the blood oxygen saturation level ($SpO_2$ as a percentage) and pulse rate (PR in bpm) for anyone on the move, such as field workers and sports enthusiasts. These devices can rapidly detect how efficiently oxygen is carried from the heart to the body extremities. Normal readings of the $SpO_2$ are between 95 to 100%, whereas a reading below 90% may indicate hypoxemia, mild respiratory diseases, and the need for supplemental oxygen. Most oximeters determine the pulse rate and $SpO_2$ within 8–10 s, display the $SpO_2$ for up to 100%, and can last up to 30 h on two AAA batteries. The results are displayed on a large digital display (organic light-emitting diode (OLED), liquid crystal display (LCD), or LED). Most of these pulse oximeters can be used both as a pediatric pulse oximeter for kids and a blood oxygen monitor for adults. Some manufacturers of these oximeters are AccuMed, BPL Medical Technologies, CVS Health, Hesley, HealthSense, iProven, mibest, NuvoMed, Roscoe Medical, TrackAid, Tomorotec, Wellue, and Zacurate.

### 5.4. Occupational Cardiovascular Diseases

The work-related risk factors for occupational cardiovascular disease are sedentary behavior, an unhealthy diet, tobacco smoke, poor air quality, lead exposure, and stress-contributing factors such as job strain and insecurity, low social contact, and a lack of decision latitude.[178,179] Symptoms of cardiovascular disease include shortness of breath, pain or discomfort in the chest or limbs, rapid or irregular heartbeat, a lack of balance or coordination, fatigue, confusion, numbness, and fainting.[180] In addition to blood tests and chest X-rays, heart health is measured by blood pressure monitoring, electrocardiography (ECG or EKG), echocardiography, the exercise stress test, and cardiac magnetic resonance imaging.[181,182]

Continuous EKG recordings for detecting an irregular heart rate or rhythm over long time periods (i.e., from 24 to 72 h) is provided by the classic Holter monitor.[183–185] The following are examples of wearables for assessing cardiovascular health.

The KardiaMobile EKG monitor by Omron is used to record medical-grade EKG data by placing fingers on a sensor pad and connecting to the Omron App.[186] There are no wires, cables or gels needed to record the electrical signals, and the monitor provides a clear view of the P, QRS, and T waves of the EKG waveforms. The Omron EKG device is cleared by the Food and Drug Administration (FDA) to detect normal heart rhythm, atrial fibrillation, bradycardia, and tachycardia in 30 s. The device can also detect normal sinus rhythm, sinus rhythm with premature ventricular contractions (PVCs), sinus rhythm with supraventricular ectopy (SVE), and sinus rhythm with Wide QRS.

iBeat has developed a smartwatch (iBeat Heart Watch) to continuously monitor the heart, blood flow, oxygen levels, and other biometrics, and notifies the wearer if an unusual event is detected using proprietary AI algorithms.[187] The iBeat Heart Watch screen is a 1.39 in. AMOLED panel with $400 \times 400$ resolution and has a 350 mAh battery. In situations where the wearer is unresponsive to the notifications, the iBeat 24 h dispatch team is contacted to send medical help through its comprehensive Heart Hero Network. An emergency button is available to contact emergency medical teams, police, firefighters, family, or friends.

The Zio system by iRhythm Technology is used to detect and diagnose irregular heart rhythms for up to 14 days.[188] The Zio Monitor is worn on the chest after prescription by a healthcare provider. The Zio Patch is a small, unobtrusive adhesive patch attached to the skin that essentially functions as an ambulatory cardiac monitor. Upon having unusual heart symptoms, the top of the Zio Patch is manually tapped. The heart symptoms are logged into a symptoms booklet or into the MyZio Symptom Tracking app. Some parameters detected by the Zio XT device are the heart rate, sinus rhythm, ventricular tachycardia, atrial fibrillation and pause, and sinus rhythm with supraventricular ectopy (SVE). After the end of the recording period, the monitor and symptoms booklet are mailed to iRhythm, which analyzes the heart data to categorize the type of arrhythmia and sends the report to the wearer's physician.[189]

A disposable health patch by IMEC is used to measure vital health signs simultaneously using its proprietary MUSEIC v3 SoC solution.[117] The IMEC chip, powered by an ARM Cortex M4 processor, has biomedical sensor readouts for measuring the heart rate using ECG, breathing rate using bioimpedance, and blood oxygen saturation through by photoplethysmography. The data preprocessing is conducted on-chip to reduce the latency and bandwidth typically associated with data transmission to the cloud or other devices. The data transfer is 100% secure because of its on-chip dedicated hardware for encryption and authentication. The IMEC health patch is suitable for ambulatory long-term monitoring of patients located at home (e.g., chronic patients or patients recovering from surgery).

## 5.5. Occupational Sun Exposure and Continuous Glucose Monitoring

Skin cancers can result from outdoor UV exposure and occupational exposure to arsenic, soot, and ionizing radiation.[190–193] The market trend indicates that skin-aware wearables and companion apps are becoming an integral part of the personalized skin care and cosmetics industry.[194] The following are examples of wearables for tracking outdoor UV exposure and managing thermal comfort.

L'Oréal has launched the My Skin Track UV device to measure the amount of UV exposure and educating the public about sun protection.[195] The device also measures the levels of other environmental aggressors such as heat, humidity, pollen, and pollution. Its battery-free wearable sensor is worn on the thumbnail, measures the UVA and UVB rays' exposure during the course of the day, and connects to its mobile app via NFC to transfer data to the user's smartphone. The wearable sensor is activated by sunlight, and hence it can last for several years. Data can be stored up to 3 months and the app shows the exposure trends over time.

QSun has built the QSun Wearable device to track the daily sunlight exposure and vitamin D production with Bluetooth low energy (BLE) communication to smartphones.[196] The QSun App provides information on UV exposure history, the daily UV index forecast, the UV index map, sunscreen reminders, and Vitamin D intake, along with options to review the progress and get personalized recommendations on sun safety based on skin type.

Embr Labs has designed the Wave bracelet as a personal thermostat for the wearer.[197] During thermal discomfort, the user activates the Wave bracelet to generate thermal or cooling waveforms through its thermoelectric heat pump targeting the thermoreceptors located in the inner region of the wrist. The local warming or cooling sensation through the device sends favorable signals to the brain to improve the overall perception of thermal discomfort.

Within the realm of wearables for the diagnosis of blood conditions, one notable example is the Dexcom technology solution for the continuous management of type 1 and type 2 diabetes.[198] The Dexcom G6 Continuous Glucose Monitoring (CGM) system uses a slim sensor inserted beneath the skin to detect blood glucose levels every 5 min. There are no finger pricks or device calibrations needed with the Dexcom G6 CGM system. A single Dexcom G6 sensor can be worn for up to 10 days. A companion app displays the data with customizable alerts about high or low blood glucose levels. Wearables in the near future will be able to relate emotions and states of mind with relevant biomarkers, muscle activity, and brain waves for informed decisions on one's current health status.[199–202]

## 6. Connected Worker Solutions for Data Management and Advanced Analytics

Workplace wearables serve as data collection devices within a central hub (i.e., connected worker solution) that monitors and manages organizational assets with minimal human intervention. Here, organizational assets refer to machinery, equipment, cameras, sensors, and even workers fitted with wearable devices that capture relevant biometrics and environmental data. **Figure 6** shows a typical architecture of a connected worker solution. Wearables and other mobile devices use appropriate device services to communicate with the edge gateway.[203] MQTT is a popular messaging protocol that uses a connect-publish-subscribe-disconnect model to transmit messages between clients and devices. MQTT is considered to be extremely lightweight because it requires a small code footprint and small network bandwidth. Other device services and networking protocols for managing and monitoring network-connected devices are Representational State Transfer (REST), OPC Unified Architecture (OPC-UA), Bluetooth Low Energy (BLE), ZigBee, and Simple Network Management Protocol (SNMP). To provide real-time insights, edge computing is the preferred choice where the data processing and decision-making nodes are closer to the data collection devices. The edge gateway is responsible for data pre-processing, data caching, edge computing, and edge analytics. The cloud hub is used for data management and storage, stream analytics, decision making, rules engine, and security and privacy.[204]

Advanced analytics with AI and ML is applied to the cloud data in the cloud hub to extract intelligent information and make predictions about asset allocations, operations, and supply chain logistics.[205] As an alternative, stream analytics can be carried out on real-time streaming data to score incoming events and provide actionable insights for future decisions.[206] Advanced AI/ML tools analyze the data from disparate sources, identify key relationships between safety events, and unravel the causal factors indicative of high-risk incidents.[207] A central dashboard or visualization studio is maintained where all the raw data, information, inferences, and insights are accessible in a comprehensible format to operators, workers, supervisors, and/or Environment, health, and safety (EH&S) personnel. The central dashboard provides real-time monitoring and management of assets, devices, field services, and event streams to the supervisor, along with bidirectional communication with the onsite operators and workers. The command station can notify remotely located supervisors about specific workers exposed to harsh chemicals, fatigue, exhaustion, and dehydration. By tracking dangerous work conditions, appropriate and timely notifications can be sent to lone workers to take recovery breaks. The central dashboard automates the tasks of scheduling (of jobs, tools, and equipment), registration and logging of operational assets, and certification and compliance checks. In addition, predictive analytics is employed to foresee equipment failure, optimize resource allocation, manage productivity trends, and identify risk-prone situations.[208] Depending on the action items, the desired outcomes of connected worker solutions are continuous monitoring, connected logistics, business process optimization, risk and cost reduction, predictive maintenance, and real-time insights about assets.

There are different options for connected worker solutions, and it is often difficult to judge the value and suitability of a connected worker solution for a given workplace. The choice of an appropriate connected worker solution depends on the proper assessment of several competing factors at the workplace, such as the desired business insights, the nature of the work environment, safety risks and hazards, platform scalability, characteristics of assets and legacy equipment, needs for predictive maintenance, network connectivity, visibility and transparency of processes, energy efficiency, skillset of existing workers and information technology (IT) staff, technology adoption barriers, information on technology validation and reliability, data ownership, privacy, security, and pricing/subscription models.

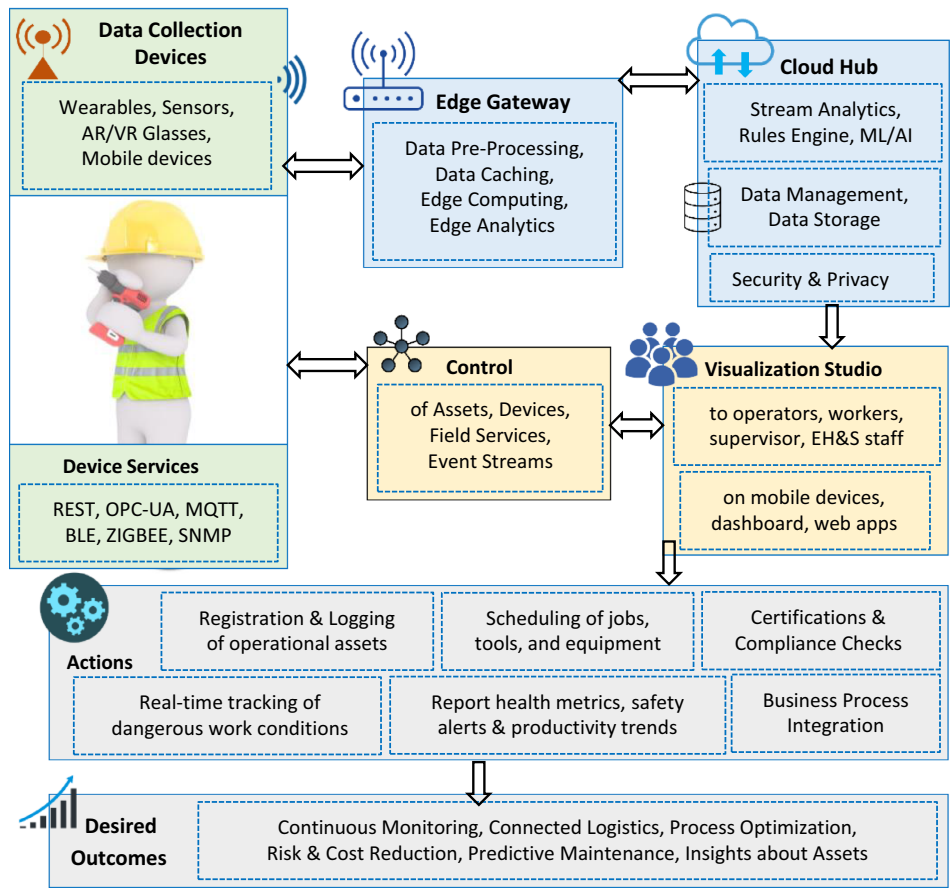

**Figure 6.** Architecture of a connected worker solution. The data collection devices use different device services to connect to the edge gateway for data preprocessing, caching, edge computing, and edge analytics. The cloud hub is responsible for data management, storage, security, and privacy. There are capabilities for streaming analytics with rule engines and AI/ML algorithms. The visualization studio displays the analyzed results on the dashboard and sends notifications to operators, supervisors, safety personnel, and workers. The strategies to control assets, devices, field services, and event streams are suggested by predictive algorithms. Finally, action items are executed to achieve the desired outcomes.

There are multiple vendors for building, hosting, managing, scaling, and customizing connected worker solutions as described below (**Table 6**).

Amazon has developed the Amazon Web Services (AWS) Internet of Things (IoT) Core platform to connect and manage a large number of devices.[209] Data management and analytics are provided by AWS cloud computing. The AWS Greengrass allows users to run AWS on edge devices using AWS Lambda functions or Docker containers for local computation and near real-time analytics. AWS IoT Device Management helps to register, regulate, and monitor IoT devices while the AWS IoT Device Defender ensures that network configurations follow the best security practices. There are services for running advanced analytics on the collected data to detect and respond to target events. The AWS IoT Things Graph generates customizable IoT configurations and sophisticated workflows based on prebuilt models of popular devices (e.g., cameras, mobile sensors) available through Amazon Simple Storage Service (S3) and Amazon Recognition.

The Cisco Kinetic platform protects clients from security risks while helping to deploy and manage operations at different scales.[210] Their four broad areas of focus are network connectivity, connectivity management, edge computing, and data control and exchange. The Cisco Kinetic platform consists of the Edge and Fog Processing Module that computes data at distributed nodes and the Data Control Module that moves the computed data to the appropriate cloud application.

Enterprise Health has created enterprise solutions with on-site clinics to integrate occupational health and safety with clinical care.[211] Enterprise Health software aggregates all workers' health information, including their immunization records, compliance, risk factors, and medical history. Its platform provides clients with a single place to document, manage, and report on occupational injury and illness. Employees have electronic access to their health records with tracking of required surveillance exams, testing, and training to maintain work certifications.

The Hitachi Lumada platform has digital solution tools and technologies that provide insights into the data from IoT

Table 6. Examples of connected worker solutions for different workplaces.

| Company | Sample products | Modular units | Measurable parameters | Connected worker applications |
|---|---|---|---|---|
| Amazon | AWS IoT Core, AWS IoT Greengrass, Amazon Services, Alexa Voice Service | IoT devices, IoT services, edge and cloud computing, pre-built IoT models, multi-layered security | Connectivity of devices and services, data analytics, visualization, security | Securely connect devices and services to the cloud, continuous monitoring |
| Cisco | Cisco Kinetic, Edge and Fog Processing, Data Control | Sensors, switches, routers, embedded networking, edge computing | Network connectivity, data control and exchange, security risks | Securely connect distributed sensors, local compute, cloud migration |
| Corvex | Smart Sensor IoT Ecosystem | IoT sensors, core devices, dashboard, sentiment meters, algorithms | Individual contributions, performance metrics, PPE compliance metrics | Promote situational awareness, frontline observation, smart PPE |
| Enterprise Health | Enterprise Health Software Solution | Health monitoring devices, medical surveillance exams, ehealth records | Health risk factors, injuries, medical history, compliance | Reporting and recordkeeping, early intervention, compliance checks |
| Hitachi | Hitachi Lumada, Digital Supply Chain, Logistics | IoT devices, HX Series controllers, 4M data streams, SaaS-based AI | Production cycle, lead time, operational efficiency, schedules | Improved logistics, predictive maintenance, remote assistance |
| Honeywell | Industrial IoT, Experion Process Knowledge System | Sensors, batch manager, controllers, gateways, edge and cloud computing | Failures of assets, connectivity, process automation, risk notifications | Centralized collaboration, early event detection, predictive analytics |
| IBM | Maximo Worker Insights, Watson Works | IoT devices such as cameras, mobile phones, wearables, other BLE devices | Occupancy, crowd density, entry/exit times, body temperature, privacy | AI-powered management of jobs, issues, anomalies, alerts |
| Intelex | EHS & Quality Management Platform | Software management tools for training, risks, audits, inspections | Details of incidents, risks, injuries, illnesses, safety training, claims | Incident report generation and tracking; safety management |
| IoTConnect | Smart Connected Worker, Smart Asset Monitoring | IoT devices, SDKs, plug-and-play connectivity, Platform as a Service | Ambient temperature, toxic gas levels, noise levels, vitals, asset health | Conditional monitoring, real-time analytics, transparency |
| Microsoft | Microsoft Azure IoT, Dynamic 365, Microsoft 365 | HoloLens 2, IoT devices, back-end services, IoT Device SDKs, Azure | Visualization tools, IoT connectivity, telemetry, storage, stream analytics | Connected worksite, work insights, fraud protection, remote assistance |
| Oracle | Oracle IoT Connected Worker Cloud | GPS trackers, wearable sensors, mobile phones, portable devices | Location, health, accidents, safety violations, KPIs | Boost worksite performance, diagnostics, regulatory compliance, |
| Rufus | WorkHero | Cloud enterprise software, Team dashboard | Worker performance metrics, custom KPIs, safety incidents | Full team visibility, dashboard insights of warehouse operations |
| Siemens | MindSphere, MindAccess, MindServices, MindConnect | IoT devices, Nano device, Industrial IoT as a service | Location, asset health, safety events, injuries, accidents, safety violations | Remote monitoring, secure cloud transfer, process optimization |

devices.[212] Hitachi works with customers to analyze issues, construct hypotheses, build prototypes of possible solutions, and execute delivery and implementation. As an example of the platform's applicability in the workplace, Hitachi's technology solutions have been used for smart manufacturing and to create next-generation factories that can handle mass customization of machine tools. This feat was accomplished by installing automated and unmanned systems and setting up IoT environments with RFID tags. The benefits were accelerated factory control cycles, better production visualization, operational efficiency, and reduced manufacturing lead times. Recently Hitachi and Microsoft formed a strategic alliance to provide next-generation digital solutions for the manufacturing and logistics sectors of Japan, North America, and Southeast Asia. This collaboration will leverage the Hitachi Digital Supply Chain, Hitachi Digital Solution for Logistics/Delivery Optimization Service, Azure IoT, HoloLens 2, and Dynamic 365 Remote Assist for huMan, Machine, Material, Method (4M) data collection, predictive maintenance, and remote assistance to workers.

The Honeywell Experion Process Knowledge System (PKS) is a distributed control platform for automating process manufacturing and managing assets or processes.[213] This system can monitor employees' fatigue levels, reduce their process trips and incidents, and help boost their productivity. Routine and abnormal work situations can be detected by its Collaborative Station and Application Control Environment, along with supervisory control solutions across multiple sites within the enterprise. Big data analytics also enables early event detection.

The IBM Maximo Asset Performance Management (APM) suite helps organizations manage their assets, extend asset lifecycles, and improve operations.[214] The dashboard aggregates data from multiple sources (such as daily work/inventory requirements, plans for scheduled work, and the progress of projects, programs and work efforts) and visualizes them in a single place. IBM's Maximo Worker Insights provides real-time insights for inventory, managing occupancy and no-go zones, tracing work orders and entry/exit times, checking adherence to privacy policies, and even monitoring employees' body temperatures. IBM's Maximo APM Predictive Maintenance Insights predicts the likelihood of future asset failures and their impact. IBM's Watson Works performs data computation and decision-making using its Watson AI models and applications.

The Intelex Environmental Health, Safety and Quality Management (EHSQ) platform has web and mobile applications to reduce risks, ensure compliance, and improve productivity.[215] Some applications integrated into this platform relate to the management of chemicals, risks, hazards, audits, inspections, employee training, and certifications. Intelex also provides the Occupational Injury and Illness software to generate reports on job-related incidents and illnesses in compliance with regulatory agencies. This enables faster and streamlined tracking and reporting of incidents with greater transparency between workers and safety managers.

IoTConnect has multiple platforms for the smart monitoring of factories, offices, assets, energy, fleets, food, air quality, and warehouses.[216] The IoTConnect Smart Connected Worker solution allows employers to remotely track their field workers to ensure a safe work environment with better regulatory compliance. It further aims to reduce unexpected accidents and the risk of errors through greater visibility of worker activities and adherence to safety procedures/policies. For example, geofences can be created to remotely track the location of workers in hazardous worksites and to send them real-time triggers when they cross a geofence. The IoTConnect platform can monitor and analyze workers' vital signs and environmental factors (i.e., noise levels, toxic gas levels, and surrounding temperature) in real time.

The Microsoft Azure IoT cloud services platform is built to connect, monitor, and manage a large number of assets.[217] The Azure IoT Hub enables device-to-cloud (and vice versa) communication and back-end services on the cloud. Microsoft's Dynamic 365 Remote Assist empowers technicians to collaborate and work efficiently across different locations using the HoloLens 2 and an Android or iOS device. This promotes virtual site inspection, workflow visualization, real-time problem solving, remote training/education, and associated cost savings. These Microsoft solutions for the workplace are gaining significant traction with clients in diverse industry sectors such as healthcare, process manufacturing, production, shipping/ports, power/utilities, agriculture, forestry, and fishery.

The Oracle Connected Worker Cloud platform provides real-time visibility of the worksite and workers.[218] Using GPS technology, this platform can track the physical location of workers in hazardous environments, monitor and alert them to possible safety violations, and analyze Key Performance Indicators (KPIs) for projects and employees. Its diagnostics analytics is used to relate worksite incidents to data from personnel, equipment, and the environment.

The Siemens MindSphere cloud operating system connects different machines, products, plants, and systems for data collection and advanced analytics.[219] The operating system is based on a Software-as-a-Service (SaaS) architecture where users connect to the cloud to analyze their data in real-time. Its open environment has APIs and services, to interface with Azure, AWS or Alibaba infrastructures. The Siemens MindConnect Nano devices enable connectivity to MindSphere for data transfer to the cloud through a secured internet connection along with edge analytics to improve operational efficiency and productivity.

## 7. Current Challenges and the Path Forward

There are some caveats in using commercial workplace wearables and connected worker solutions for the safety health, and productivity of workers.[2,4] a) Only a few workplace technologies have gone through rigorous field studies, extensive validation against standards, and approval processes by related agencies. With a lack of sufficient peer-reviewed data, it is difficult to assess the safety and efficacy of most workplace technologies in the market. b) There are heterogeneity and compatibility issues among the available workplace technologies as developers are reluctant to adhere to common evaluation practices. For example, certain wearable devices may only be compatible with selected operation systems (e.g., Google's Wear OS, Apple's Watch OS, Garmin OS, WebOS, Samsung's Tizen, MediaTek's LinkIt, Linux-derivative etc.), which may be a barrier for communicating with other existing devices running on alternate operating systems. Similarly, remote worker health monitoring may require multiple heterogeneous sensors (e.g., motion sensor, electrocardiogram, heart rate monitor, blood oxygen monitor, etc.) to get an overall understanding of health status, but these heterogeneous sensors may not be supported by the same operating system, SDK, or data transfer protocols. Moreover, different providers may terminate or relinquish the support of their existing wearables, apps, and services in this rapidly evolving and competitive marketplace. This creates a challenge for actual users on how to choose and invest in cross-technology platforms that can seamlessly communicate across multiple products. c) There is a general lack of consensus on how to integrate raw data from multiple sources (in situ and in real time) at the level of the individual while performing different operational tasks. Inculcating a proper data-driven culture and analytical acumen within the IT staff takes time and effort. As such, most employers run the risk of ceaselessly collecting data across disparate sources and making investments in data warehousing and business intelligence tools. d) The extensive use of workplace technologies may lead to device fatigue, over-dependency on such technology, and helplessness in the event of technology outages. An excessive monitoring of workers in the workplace may result in the Hawthorne effect where a user's behavior can be altered due to the awareness of being observed. Workers equipped with wearables may lose opportunities to interact with colleagues and their autonomy in performing tasks. e) As wearables have the potential to collect private information (e.g., work hours, breaks clocked, sleep hours, interests, habits, social circles, and even health records), there is little clarity on who actually owns the user data or how they intend to use the data.[32] The collected information could facilitate bias, prejudice and discriminatory behavior in the workplace. Wearables and mobile computing devices are also prone to security lapses and data breaches.[203] f) The general acceptability of data analysis tools is a realistic challenge for the workforce.[32] It is difficult to trust the judgments and scoring systems of AI/ML algorithms, particularly in matters concerning people's health and safety. One survey indicated that company executives have made major decisions based on their "own intuition or experience" or the "advice and experience of others internally" rather than "data and analysis."[220] Another survey of 480 organizations found that merely 4% of them were using predictive analytics to forecast outcomes, 86% of them focused primarily on operational reporting and only 10% of these organizations used some sort of statistical model and root-cause analysis.[221]

The outlook of commercial workplace wearables and connected worker solutions is very promising, considering the power of these technologies to aggregate data from heterogeneous sensors in a continuous manner and apply predictive analytics to identify organizational risks and workflow bottlenecks for a myriad of industrial applications. To thrive in this competitive and rapidly evolving market, device manufacturers often need to prioritize their efforts towards a growing customer base with recurring revenues and profits. Toward this goal, there are several directions to build a successful business model—from understanding the technology adoption paradigms to large-scale validation studies of proposed methods. a) Conceptual models (e.g., Theory of Reasoned Action (TRA), Technology Acceptance Model (TAM), and Unified Theory of Acceptance and Use of Technology (UTAUT)) can help identify the influencing factors for technology adoption, such as behavioral intention, trust, risks, ethics, perceived privacy risk, perceived usefulness, perceived ease of use, perceived vulnerability, social influence, and demographics. b) Large-scale studies conducted by independent third parties and end-user companies can help to develop validation and reliability models, which in turn could attract future clients and investments. Several companies in this domain have active collaborations with end-user companies to generate large data at the worksite, validate their AI/ML models, and identify new applications. c) Financial incentives, academic grants, and crowdsourcing research can bring in researchers, scholars, and students from universities to validate and improve their end-to-end data pipeline and generate a body of scientific publications, white papers, and case studies. This approach is being pursued by leading cloud service providers, including IBM, Microsoft, Google, Apple, Amazon, Oracle, Cisco, Siemens, and Honeywell. Creating open-source hardware and software platforms with seamless process integration can be a successful route to spur community science, and has been proven by companies such as Arduino, Adafruit Industries, Raspberry Pi, MakerBot, and Sparkfun. d) Obtaining regulatory clearance (e.g., FDA 510(k) clearance) and safety compliance certifications can boost consumer confidence, open up new markets, connect with reseller networks, and expand the customer base.

As a path forward to invest in and explore the benefits of workplace technologies, there are some points to consider, particularly for the corporate leadership, safety managers, and supervisors. a) A cost-benefit analysis needs to be carried out to understand the value proposition of introducing new technologies with respect to existing manpower, skillsets and legacy assets.[222] The benefits could be in terms of continuous and unobtrusive monitoring, connected logistics, process optimization, predictive maintenance, real-time insights about assets, cost reduction, increased revenue, and higher productivity.[9,206] b) Worker education is necessary to explain the rights of workers, address their apprehensions, and gain their acceptance.[32] Appropriate departments need to regulate data collection, analytics, control compliance, and security. In an unprecedented move to promote data transparency and privacy, Apple has recently started to disclose to its customers the types of personal data collected by the apps installed in its iOS devices. Apple will also release an antitracking feature where apps will have to ask for permission prior to tracking the user's activities.[223] c) It is advisable to generate baseline curves for individual workers in realistic workplaces having limited power budgets and cloud connectivity.[27] It is critical to recruit committed users who can provide real-time feedback and revalidation with regard to their mental and corporeal status, behavioral preferences, technology adherence, and overall benefits.[224] Doing so will ensure that ground truth data are accurate, credible, scalable, and of high quality and granularity.[203] d) Finally, only the right conclusions should be drawn from data analytics and model tools.[225,226] It is important to tell the story behind the data to all stakeholders in simple language, while acknowledging the fact that even the most advanced algorithms cannot capture all possible real-world scenarios indicative of occupational risks and hazards.[9–12,227,228]

## 8. Conclusion

Protecting and improving the safety, health, and productivity of workers is paramount for companies as they face post-pandemic challenges of acute shortage of skilled labor, strains on supply chain management, global inflation, and rising medical costs. In this regard, intelligent systems (deployable sensors and analytics) play a critical role to facilitate TWH through continuous monitoring, management, and prediction of workplace risks and organizational assets. In this work, we reviewed the recent trends in commercial wearable technologies and connected worker solutions being applied to different work settings to promote ergonomics, situational awareness, injury risk management, efficient workflow, and healthy behavioral and cognitive habits. While a majority of devices monitor human performance (e.g., biomechanical functions, physical activity, and/or physiological signals), new intelligent systems are being introduced to actively monitor and manage mental health (e.g., stress, emotions, states of mind) using brain wave sensing, biofeedback and human-in-the-loop models. There is ample scope for device manufacturers to expand their customer base, considering the predictive power of these intelligent systems to identify a variety of occupational risk scenarios for timely intervention. To enable large-scale technology adoption at the workplace, it is important for device manufacturers to validate the reliability their products by independent third-party testing. At the same time, technology consumers need to identify the true value proposition of these technologies with respect to legacy assets. Toward this end, our work attempts to bridge the knowledge gap between device manufacturers and technology consumers by providing a comprehensive review of recent workplace technologies.

## Acknowledgements

V.P. and A.C. contributed equally to the work. The authors are grateful to the companies for taking the time to discuss their technologies with them and providing permission to reproduce their technology figures in this publication. This work was partially supported by the USA. National Science Foundation (NSF IDBR-1556370) and USA Department of Agriculture (2020-67021-31964) to S.P.

## Conflict of Interest

The authors declare no conflict of interest.

## Data Availability Statement

Research data are not shared.

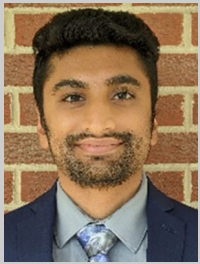

**Vishal Patel** received his Bachelor's degree in electrical engineering and Masters of Business Administration (MBA) from Iowa State University, in 2021. He is currently working in the medical industry as a field service engineer where his interests lie in the areas of robotics and power systems.

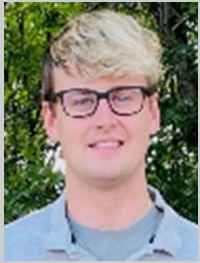

**Austin Chesmore** received his Bachelor of Science degree in electrical engineering from Iowa State University, in 2020. He is now working in the electric vehicle (EV) industry as an electrical/embedded software engineer where his interests are in electrical system design, control systems, robotics, EV design, and human–computer interaction.

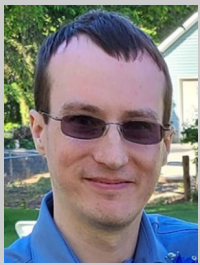

**Christopher M. Legner** received his Ph.D. in electrical engineering from Iowa State University, in 2020. He is now working in the agricultural industry as an automation specialist where his interests lie in areas of robotics, automated sample preparation, instrument design, and bioinformatics.

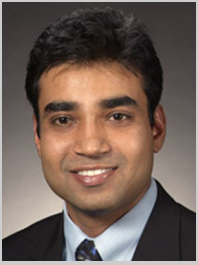 **Santosh Pandey** is an associate professor in the Department of Electrical and Computer Engineering at Iowa State University. His laboratory focusses on building new devices and systems that automate the experimental methods in life sciences. The lab provides hands-on research experience to a number of undergraduate and graduate students every year on a wide range of topics from device fabrication and testing to data-driven model development and validation.